\documentclass[showpacs,prb,preprintnumbers,amsmath,amssymb,twocolumn]{revtex4}

\usepackage{graphicx}
\usepackage{dcolumn}
\usepackage{bm}
\usepackage{bbm}
\usepackage{amsmath}
\graphicspath{{plots/}}

\begin{document}

\title{Multiband Effects on Superconducting Instabilities Driven by Electron-Electron Interactions}

\author{Stefan Uebelacker}
\email{uebelacker@physik.rwth-aachen.de}
\author{Carsten Honerkamp}
\affiliation{%
Institute for Theoretical Solid State Physics, RWTH Aachen University, D-52056 Aachen 
\\ and JARA - FIT Fundamentals of Future Information Technology
}%

\date{November 4, 2011}

\begin{abstract}
We explore multiband effects on $d$-wave superconducting instabilities driven by electron-electron interactions. Our models on the two-dimensional square lattice consist of a main band with an extended Fermi surface and predominant weight from $d_{x^2-y^2}$ orbitals, whose orbital character is influenced by the admixture of other energetically neighbored orbitals. Using a functional renormalization group description of the superconducting instabilities of the system and different levels of approximations, we study how the energy scale for pairing and hence the critical temperature is affected by the band structure. We find that a reduction of orbital admixture as a function of the orbital energies can cause a $T_c$ enhancement although the Fermi surface becomes more curved and hence less favorable for antiferromagnetic spin fluctuations. While our study does not allow a quantitative understanding of  the $T_c$ differences in realistic high-$T_c$ cuprate systems, it may reveal an underlying mechanism contributing to the actual material trends.
\end{abstract}

\maketitle

\section{\label{introduction}Introduction}
With the advent of the new iron superconductors,\cite{pnictidereview} there is now a second class of superconducting materials  with high $T_c$'s besides the layered cuprates. As opposed to the cuprates, the iron pnictide superconductors have more than one electron band at the Fermi level, and the present theoretical works to understand these materials clearly indicate that this fact makes the systems far more complex. In particular, theorists have undertaken efforts to relate the observed differences in the transition temperatures and in superconducting gap structures through the families of iron superconductors to details in the crystal and hence electronic structure.\cite{kurokipnictides,valenti,platt,ikeda} Similar ideas were then applied to the cuprates as well.\cite{sakakibara} These works tackle the full complexity of many-orbital problem with approximate many-body techniques. The results are interesting and promising, as they show the existence of various tuning parameters for $T_c$. On the other hand, we feel that the  problem of correlation-driven superconductivity should also be approached from a constructive point of view using simpler models, by asking what changes occur if one takes a separated Fermi surface with a superconducting instability, and adds to it the orbital character of the band, or allows other bands to come close in energy. The main question is whether one can find explainable trends that can be used as guide lines in a search for higher $T_c$'s or other desired properties. This is one of the motivations for the study described below. 

Another direct motivation for considering electron-electron-interaction-driven pairing in multiband models is the $T_c$ trend in the high-$T_c$ cuprates. As pointed out ten years ago by O.K. Andersen's group,\cite{pavarini} there appears to be a positive correlation between the experimental $T_c$'s and the theoretically derived second-nearest-neighbor hopping parameter $t'$ or a related parameter $r$. Here, the higher $T_c$'s occur for a rounder Fermi surface, i.e., for larger $t'$ and larger $r$. This contradicts at least the naive expectations in a spin-fluctuation-induced pairing scenario, where a smaller amount of nesting leads to a weaker pairing interactions, and hence a round Fermi surface with larger $t'$ should, at least over some parameter range, have a smaller $T_c$.  This thinking is purely based on the geometry of the Fermi surface (plus van Hove singularities nearby) and does not include any information on the orbital content of the band at the Fermi level. Also numerical studies on the one-band Hubbard model using the dynamical cluster approach\cite{maierdca} and density matrix renormalization group calculations on $t$-$J$ ladders\cite{white} have shown the trend that a larger $t'$ leads to smaller $T_c$, in contradiction with the above mentioned findings. Only in $t$-$t'$-$J$,  comprehensive Quantum Monte Carlo (QMC) studies by Spanu \textit{et al.} \cite{spanu} showed a slight enhancement of pairing  at optimal doping with nonzero $t'$, however, to a smaller extent than in earlier variational QMC studies by Shih \textit{et al.}\cite{shih} Hence in the one-band model, both at weak and strong coupling, the theoretical expectations are inconclusive and certainly not fully consistent with the empirical trend.  The simple question now is whether the theoretical picture is altered when orbital information of the multiband case is included. Therefore we revisit this problem in simple multiband models for the band structure of the cuprates.

The change of the next-to-nearest-neighbor hopping parameter $t'$ is caused by changes in the multiorbital electronic structure of the cuprates. In downfolded four-orbital models for the cuprates,\cite{pavarini,hansmann} its increase is related to a lowering in energy of the so-called axial orbital toward the Cu $3d_{x^2-y^2}$-level. The axial orbital is basically a linear antibonding combination of the local Cu $4s$ and the surrounding oxygen $2p_z$ states. It can be decreased by reducing the overlap between these two orbitals, which happens if the oxygens move further out of plane. In this way, crystal and electronic structures are correlated, and the hope is to relate the structural differences between different cuprates to the differences in their superconducting properties.

The material trend pointed out by Pavarini \textit{et al.}\cite{pavarini} also spurred exciting suggestions to produce band structures and Fermi surfaces with even higher $r$ parameters in LaNiO$_3$/LaAlO$_3$-heterostructures.\cite{hansmann} For such systems, taking over the trend from the cuprates would result in $T_c$'s above 100$K$. Therefore, a more detailed understanding of the relation between the low-lying electronic structure and the superconducting transition temperature becomes an important question in the field of tailored transition metal oxide systems.

The relation of finer differences in the multiband electronic structure to the pairing strength for different cuprates has been addressed theoretically by at least two works. Kent \textit{et al.}\cite{kent} studied downfolded $dpp$ three-band Hubbard models using the dynamical cluster approach with QMC impurity solver. The authors found a very strong sensitivity of the resulting $T_c$ on choice of the downfolding technique and the localization of the Wannier functions. Small longer-ranged hopping had a marked influence on the results.  We take this as an indication that a direct parameter-free theoretical approach with a nonperturbative (cluster) many-body technique is still too challenging, and that a qualitative understanding of how the different building blocks of the model affect the resulting $T_c$'s would be very useful.   
  
A theoretical study that was successful in obtaining a significant difference in pairing strength between La-based and Hg-based cuprates in the right direction came from the group around Kuroki, Arita, and Aoki.\cite{sakakibara} These authors considered two- and three-orbital models obtained using maximally localized Wannier orbitals to represent the low-lying density functional theory (DFT) band structure. The models were then treated by the fluctuation-exchange approximation (FLEX), which yields effective coupling strengths for $d$-wave pairing. Upon changing the oxygen height $h_{\text{O}}$ with respect to the Cu plane, the band structure was altered continuously from a situation corresponding to La$_2$CuO$_4$ to roughly that of HgBa$_2$CuO$_4$. The parameter in the two- or three-orbital models, that responded most to this structural change, was the difference $\Delta E$ of the onsite kinetic energy between the two Cu $3d_{x^2-y^2}$ ($d_{x^2-y^2}$)- and Cu $3d_{3z^2-r^2}$($d_{z^2}$)-dominated Wannier states. If a third Cu $4s$($s$) orbital was included into the model, its energy difference to the $d_{z^2}$ was found to be roughly constant along this patch, while in the two-orbital model, this change was effectively absorbed into the model parameters. In both models, the coupling strength for $d$-wave pairing was found to increase when $h_{\text{O}}$ or $\Delta E$ were increased, i.e., when the axial $s$ level moved closer to the $d_{x^2-y^2}$-level from above, and the  $d_{z^2}$ moved further down below the $d_{x^2-y^2}$-level. Along this patch the Fermi surface became more rounded. This study shows, that differences in pairing strength can indeed be related to changes in two- or three-band model parameters. The present study was motivated by this work. Our goal was to acquire a clearer understanding why these model parameters changes actually the pairing strength in a $d$-wave pairing situation, and if one can identify a simple mechanism behind the observed trends. This might be useful for prescribing other band structures that should have high transition temperatures. Below we will show that in the weak-coupling picture, there is actually a parameter window, where the detrimental effect of orbital admixture to the $d_{x^2-y^2}$-like conduction band is reduced more strongly than the other negative factor of Fermi surface rounding increases. This can lead to an increase in the pairing scale as function of the relevant parameters.

This paper is organized as follows. First, we describe the underlying models and the functional renormalization group (fRG) method used for the study of these models. Then, we apply the fRG to the two-band model and find a first trend. This trend can then be understood more deeply by using the simplified two-patch model. Finally, we then use the three-band description that allows us to qualitatively reproduce the trends seen in the FLEX calculation\cite{sakakibara} and to establish a simple picture. 
 
\section{Description of the models and the scheme}
In this paper, we first study a two-orbital model that can be thought to arise from one $s$-like and one planar $d_{x^2-y^2}$-like orbital on the two-dimensional square lattice.\cite{andersen} All other relevant orbitals, e.g., oxygen $p$ states, should be considered as included in these effective orbitals. The free Hamiltonian reads
\begin{equation}
\label{eqtwoband }
H = \sum_{\vec{k},\sigma} 
\begin{pmatrix}  c^s_{\vec{k},\sigma}   \\  c^d_{\vec{k},\sigma}   \end{pmatrix}^\dagger
\begin{pmatrix}
    \epsilon_s (\vec{k}) + \Delta E_{sd} -\mu &  v(\vec{k})   \\
v(\vec{k})        &  \epsilon_d (\vec{k}) -\mu
\end{pmatrix}
\begin{pmatrix}  c^s_{\vec{k},\sigma}   \\  c^d_{\vec{k},\sigma} \end{pmatrix} 
\end{equation} 
with the nearest-neighbor hopping dispersions $ \epsilon_{s/d} (\vec{k}) = -2t_{s/d} ( \cos k_x + \cos k_y)$, the hybridization term $v(\vec{k}) = -2t_{sd}  ( \cos k_x - \cos k_y)$ and the chemical potential $\mu$. Here, $t_{s/d}$ denotes the hoppings within the $s$ or $d_{x^2-y^2}$ orbital respectively, $t_{sd}$ is the hopping between the two orbitals, $\vec{k}$ is the momentum vector in the Brillouin zone and $\sigma$ denotes the spin projection. $c^{{s/d}(\dagger)}_{\vec{k},\sigma}$ is the annihilator (creator) of a single-particle excitations in the $s$ or $d_{x^2-y^2}$ orbital respectively. The symmetry of $v(\vec{k})$ is due to the symmetry of the orbitals that hybridize. It is zero in the Brillouin zone diagonal and strongest at $(\pi,0)$, $(0,\pi)$. The lattice constant has been set to unity.

The matrix in orbital space can be diagonalized by a unitary transformation, leading to two bands. The new operators $a^{b(\dagger)}_{\vec{k},\sigma}$ that annihilate and create the single-particle excitations in band $b$ are given by 
\begin{align}
\label{eqtrafo}
a^b_{\vec{k},\sigma}=& \sum_{o} u^*_{bo}(\vec{k}) \, c^o_{\vec{k},\sigma} \,  ,\notag \\
c^o_{\vec{k},\sigma}=& \sum_{b} u_{bo}(\vec{k}) \, a^b_{\vec{k},\sigma} \, ,
\end{align}
where the sum in the first line runs over the $d_{x^2-y^2}$ and the $s$ orbital and the sum in the second line runs over the two corresponding bands. The matrix element of the transformation at a given momentum vector $\vec{k}$ is denoted by $u_{bo}(\vec{k})$. In the case of the three-band-model given in Eq. (\ref{3band-hamiltonian}), the transformation additionally includes the contribution from the $d_{z^2}$ orbital.
 
\begin{figure}

\includegraphics[scale=.65]{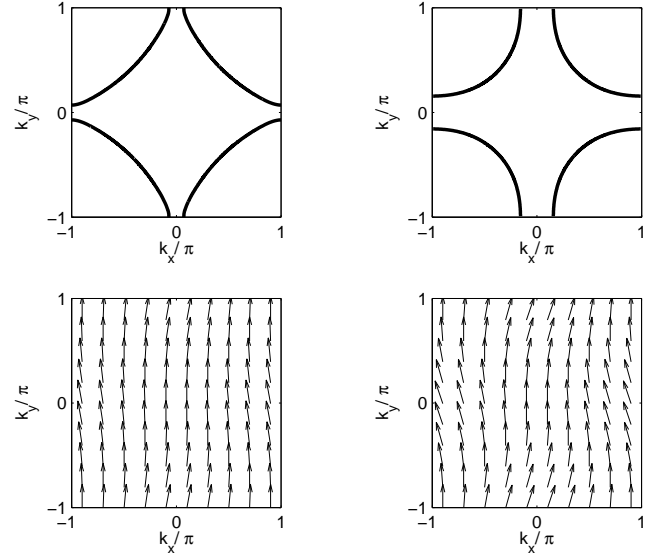}
\caption{Left: Fermi surface of the two-band model for large $\Delta E_{sd}=12 eV$; $t_d=0.45 eV$, $t_s=0.5 eV$, $t_{sd}=0.6 eV$ (upper plot), and the orbital weights of the lower $d_{x^2-y^2}$-like band through the Brillouin zone (lower plot). The vertical component of the vector arrows represents the $d_{x^2-y^2}$ orbital admixture, and the horizontal component denotes the $s$ orbital admixture. The hybridization is strongest near the $(\pi , 0)$-  and $(0,\pi)$-points. The filling is fixed to $\langle n \rangle =0.84$ per site. Right: The same for smaller $\Delta E_{sd}=6 eV$.}
\label{2bandplots} 
\end{figure}

For the creation operators, we have to use the adjoint equations. If the hybridization is not too strong, we can talk about a $s$-dominated and a $d_{x^2-y^2}$-dominated band. 
We now consider the situation where $\Delta E_{sd} $ is rather large and positive, and only the $d_{x^2-y^2}$-dominated band has a Fermi surface.  For chemical potential $\mu=0$ and large $\Delta E_{sd}$, the Fermi surface is a perfectly nested square, while if we decrease  $\Delta E_{sd}$, the Fermi surface gets rounded by the hybridization term. For the lower band around the Fermi level, we have the dispersion
\begin{align}
\label{ }
&E_c (\vec{k} ) = \frac{\epsilon_s (\vec{k})+\Delta E_{sd}+ \epsilon_d (\vec{k}) }{2} -\mu\notag\\
&- \sqrt{ \left[ \frac{\epsilon_s (\vec{k})+\Delta E_{sd}-
 \epsilon_d (\vec{k}) }{2} \right]^2 + v^2(\vec{k})} \; .
\end{align}
The Fermi surface opens up at the van Hove points and becomes similar on shape as the Fermi surfaces observed in high-$T_c$ cuprates. In Fig. \ref{2bandplots} we show two  examples. In these plots, the maximal admixture of the $s$ orbital to the band with Fermi surface is only about 10$\%$, but below we will see that already this has a measurable effect on the critical scale.

Now, we include interaction in the form of intraorbital and interorbital repulsions, $U_{s/d}=U$ and $U'$ as well as a Hund's rule coupling $J_H$ and pair hopping term $J_P$. We write the interaction Hamiltonian as
\begin{align}
\label{hinteraction}
&H_I =  U \sum_{i,o} n_{i,\uparrow}^{o}n_{i,\downarrow}^{o} +\frac{U'}{2} \sum_{i,\sigma, \sigma',\atop o \ne o'} n_{i,\sigma}^{o}n_{i,\sigma'}^{o'} \notag\\
+&\frac{J_H}{2}  \sum_{i,\sigma, \sigma',\atop o \ne o'}  c^{o \dagger}_{i,\sigma} c^{o' \dagger }_{i,\sigma'} c^{o}_{i,\sigma'} c^{o'}_{i,\sigma} +\frac{J_P}{2}  \sum_{i,\sigma, \sigma',\atop o \ne o'}  c^{o \dagger}_{i,\sigma} c^{o \dagger}_{i,\sigma'} c^{o'}_{i,\sigma'} c^{o'}_{i,\sigma} \, . 
\end{align}
The indices $i$ and $o,o'=\{s,d\}$ denote the lattice sites and orbitals respectively and $n_{i,\sigma}^o=c_{i,\sigma}^{o \dagger} c_{i,\sigma}^{o}$.
More generally we can write this interaction in wave vector space as 
\begin{eqnarray}
\label{trafoint}
H_I &=& \frac{1}{2 \, \mathcal{N}}�\sum_{\vec{k}_1, \vec{k}_2, \vec{k}_3, \sigma, \sigma' \atop o_1, o_2, o_3, o_4} 
V_{o_1,o_2,o_3,o_4} (\vec{k}_1,\vec{k}_2,\vec{k}_3)
\nonumber \\ && \times \,  c^{o_3\dagger}_{\vec{k}_3,\sigma}
c^{o_4\dagger}_{\vec{k}_4,\sigma'} c^{o_2}_{\vec{k}_2,\sigma'} c^{o_1}_{\vec{k}_1,\sigma} \, , 
\end{eqnarray} 
where $\vec{k}_4$ is fixed by momentum conservation on the lattice and $\mathcal{N} $ is the number of unit cells. $V_{o_1,o_2,o_3,o_4}(\vec{k}_1,\vec{k}_2,\vec{k}_3)$ has to be chosen appropriately, but is wave vector independent for local interactions (and has otherwise some simple $\vec{k}$ variation).
We can now write this interaction using the band operators $a_{\vec{k},\sigma}^{b (\dagger)}$. This gives
\begin{eqnarray}
\label{omakeup}
H_I &=& \frac{1}{2 \, \mathcal{N}}�\sum_{\vec{k}_1, \vec{k}_2, \vec{k}_3, \sigma, \sigma' \atop o_1, o_2, o_3, o_4} 
V_{o_1,o_2,o_3,o_4} (\vec{k}_1,\vec{k}_2,\vec{k}_3) \nonumber \\[1mm] && \times \, \sum_{b_1,b_2,b_3,b_4}
u_{o_1b_1}(\vec{k_1}) u_{o_2b_2}(\vec{k_2})u^*_{o_3b_3}(\vec{k_3})u^*_{o_4b_4}(\vec{k_4})
\nonumber \\[1mm] && \times \,
a^{b_3\dagger}_{\vec{k}_3,\sigma} a^{b_4\dagger}_{\vec{k}_4,\sigma'} a^{b_2}_{\vec{k}_2,\sigma'} a^{b_1}_{\vec{k}_1,\sigma} \nonumber \\
&=& \frac{1}{2 \, \mathcal{N}}�\sum_{\vec{k}_1, \vec{k}_2, \vec{k}_3, \sigma, \sigma' \atop b_1, b_2,b_3, b_4} V_{b_1, b_2, b_3, b_4} (\vec{k}_1, \vec{k}_2, \vec{k}_3) \nonumber \\ && \times \, 
a^{b_3\dagger}_{\vec{k}_3,\sigma} a^{b_4\dagger}_{\vec{k}_4,\sigma'} a^{b_2}_{\vec{k}_2,\sigma'} a^{b_1}_{\vec{k}_1,\sigma} 
\, . 
\end{eqnarray} 
The factor $u_{o_1b_1}(\vec{k_1}) u_{o_2b_2}(\vec{k_2})u^*_{o_3b_3}(\vec{k_3})u^*_{o_4b_4}(\vec{k_4})$ containing the orbital content is also referred to as ``orbital makeup.''\cite{maier} 

The orbital content is the new aspect in comparison with one-band models. In the multiband case, already the bare interactions in band language are wave vector dependent. 
In context with unconventional superconductivity, this raises interesting questions, in particular, whether the bare interaction already contains an attractive component in some symmetry channel, or if the generation of such an attraction (that in one-band models is usually accomplished by particle-hole-fluctuations) is somehow influenced by this extra structure. 
In the Sec. \ref{sect-twopatch} on the two-patch model, we will give more details which interaction processes are increased by that and which are reduced. We will see that under quite general conditions, the orbital content does not help $d$-wave superconductivity. 

Below we will also study a three-band model, which now also contains an orbital below the Fermi level, with the symmetry of a $d_{z^2}$-like orbital. For this model, we use parameters obtained with the Wien2Wannier scheme.\cite{wien2wannier} We drop all hoppings in the third direction, in  order to keep the model two dimensional.
The free part of the Hamiltonian is then given by 

\begin{equation}
\label{3band-hamiltonian}
H = \sum_{\vec{k},\sigma}
\begin{pmatrix}  c^s_{\vec{k},\sigma}   \\  c^d_{\vec{k},\sigma} \\ c^z_{\vec{k},\sigma} &  \end{pmatrix}^\dagger
\begin{pmatrix}
\epsilon_s (\vec{k})  &  h_{sd}(\vec{k}) & h_{sz} (\vec{k}) \\
h_{sd}(\vec{k}) & \epsilon_d (\vec{k})  & h_{dz} (\vec{k})  \\
h_{sz}(\vec{k}) & h_{dz}(\vec{k}) & \epsilon_z (\vec{k})	  
\end{pmatrix}
\begin{pmatrix}  c^s_{\vec{k},\sigma}   \\  c^d_{\vec{k},\sigma} \\ c^z_{\vec{k},\sigma} &  \end{pmatrix}
\end{equation} 
with the diagonal terms




\begin{align}
\label{hoppings1}
&\epsilon_{o} (\vec{k})=\sum_{m_x,m_y} t^{o}_{m_x,m_y} e^{i(m_x k_x+m_y k_y)} -\mu
\end{align} 
and the hybridization terms
\begin{align}
\label{hoppings2}
&h_{o_1,o_2} (\vec{k})=\sum_{m_x,m_y} t^{o_1,o_2}_{m_x,m_y} e^{i(m_x k_x+m_y k_y)} \, ,
\end{align} 
where the coefficients $t^{o}_{m_x,m_y}$ and $t^{o_1,o_2}_{m_x,m_y}$, with $o$, $o_1$, $o_2$ denoting the respective orbitals, are the parameters derived from the \textit{ab-initio} calculation.\cite{wien2wannier} The indices $s$, $d$, and $z$ are used to label the $s$ , $d_{x^2-y^2}$, and $d_{z^2}$ orbitals respectively, to shorten the notation. We include hoppings between sites on the two-dimensional square lattice with a maximum distance of five lattice constants in each lattice direction, that is the sum in Eqs. (\ref{hoppings1}) and (\ref{hoppings2}) runs over all integer values $m_x$,$m_y$ from $-5$ to $+5$. 

\begin{figure}
\begin{center}
\includegraphics[scale=0.25]{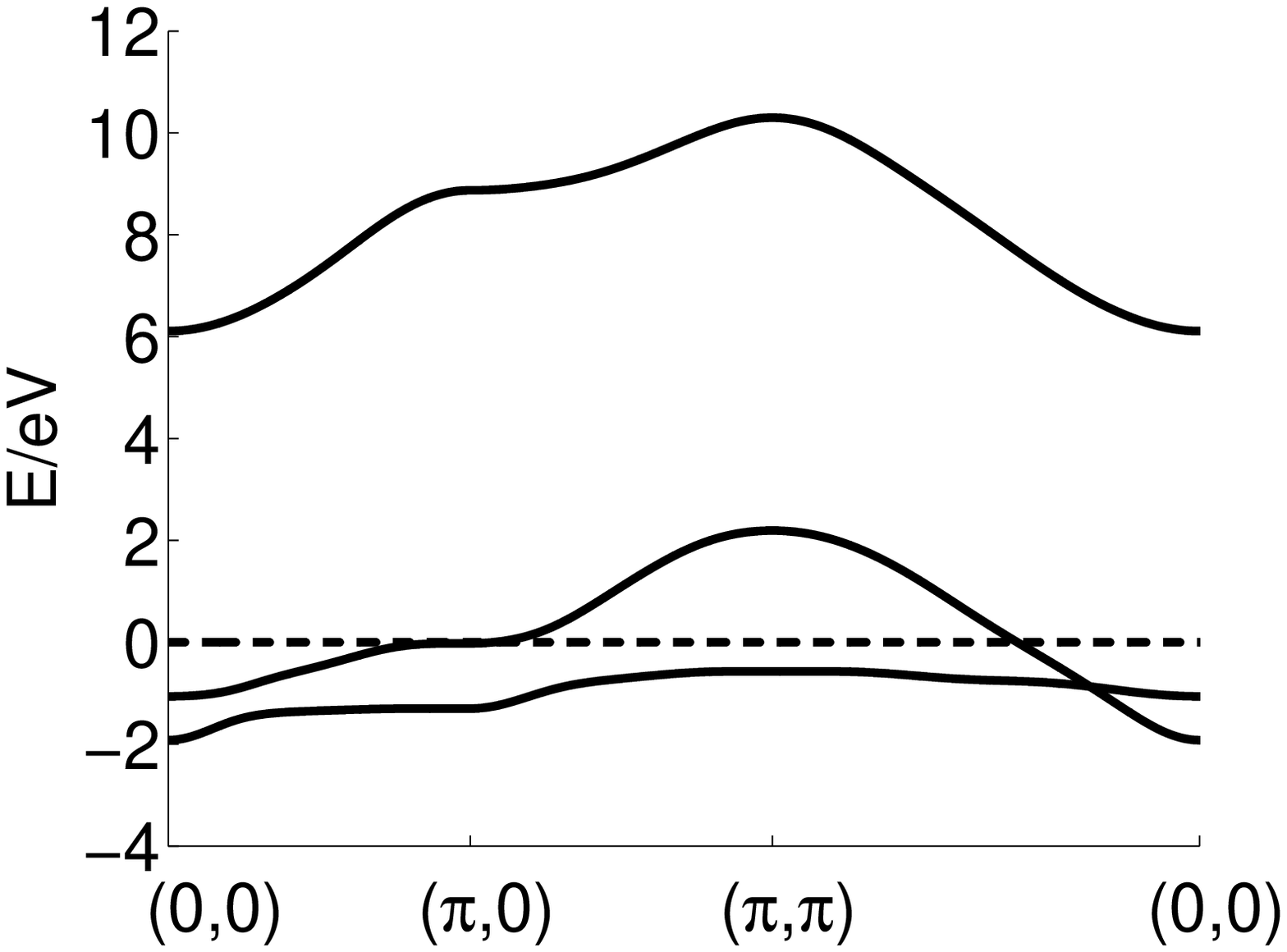}
\includegraphics[scale=0.25]{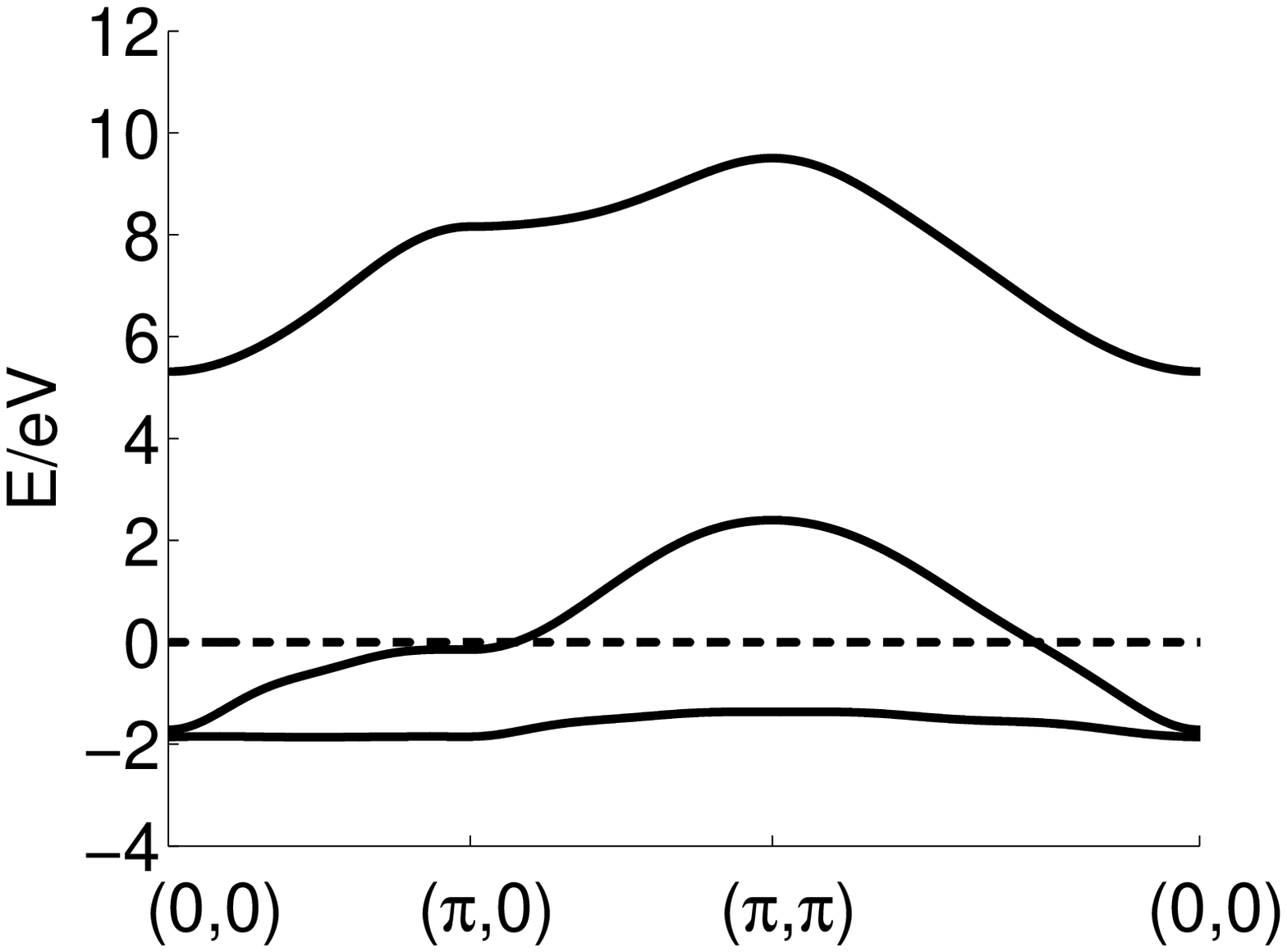}
\end{center}
\caption{Left: band structure of the three-band model with $s$-,  $d_{x^2-y^2}$- and $d_{z^2}$-like orbitals, with parameters corresponding to La$_2$CuO$_4$. Right: same band structure for larger $\Delta E=t^d_{0,0}-t^z_{0,0}$ but same $t^{s}_{0,0}-t^z_{0,0}$, roughly corresponding to HgBa$_2$CuO$_4$ according to Wien2Wannier.\cite{wien2wannier} Filling: $\langle n \rangle=2.84$.}
\label{spaghettiplot} 
\end{figure}

The resulting band structures are plotted in Fig. \ref{spaghettiplot}. Now we have one $s$-dominated band above the Fermi level and one $d_{z^2}$-dominated band below the Fermi level. The band at the Fermi level is primarily of $d_{x^2-y^2}$-character.

When applying the fRG technique described below, we only integrate out the $d_{x^2-y^2}$-like conduction band with the Fermi surface, as described in Sec. \ref{rg}. We use a basis in which the quadratic part of the Hamiltonian is diagonal and thus have to transform the interacting part according to Eq. (\ref{trafoint}). Thus the influence from the $d_{z^2}$ and $s$ orbital is then included in the perturbation of the dispersion relation and the transformation of the interaction, i.e. the orbital mixing of Eq. (\ref{omakeup}).

Again we use interactions of intraorbital, interorbital, Hund's rule and pair hopping type according to Eq. (\ref{hinteraction}), where now the sums $o$ and $o'$ go over all three orbitals $o=\{s,d,z\}$. In order to keep the analysis simple, we refrain from using orbital dependent interaction parameters.

\section{Functional RG treatment for the conduction band and additional second-order correction}
\label{rg}
In this paper, the relevant situation is when the $d_{x^2-y^2}$-like band has a Fermi surface and the other $s$-like and $d_{z^2}$-like bands are energetically separated from the Fermi surface. Then, if we are interested in the interaction effects at low temperatures and set up a perturbation expansion in the bare interactions, we can expect that virtual excitations into the bands away from the Fermi level only play a smaller role, as they generally lead to larger energy denominators in the corresponding diagrams.
The more important point where the multiband character comes in may be the orbital content in Eq. (\ref{omakeup}) that makes the bare interactions projected onto the lower band wave vector dependent. In the following we will be concerned with the question how these extra factors in front of the interaction constants affect the critical scales for $d$-wave pairing.
 
The next step is to sum the perturbation series in the interaction, i.e., to integrate out the band that forms the Fermi surface. This is done using the fRG $N$-patch approach, detailed, e.g., in Ref. \onlinecite{hsfr} and \onlinecite{rg}. This scheme takes into account the fermionic single-particle excitations step by step, with a decreasing infrared cutoff $\lambda$. 

This leads to flow of the effective interactions for quasiparticles below this scale that contain the multiple scattering processes with higher-energy intermediate particles. 
The present approach makes three simplifying approximations discussed in previous works.\cite{hsfr,rg} First self-energy is neglected in the flow. We expect the self-energy to become important only when the effective interactions become large, and we stop the flow when this occurs. 
Then, second, the one-particle-irreducible six-point vertex is ignored. Such a term is zero in the bare interaction, but will get generated. Its feedback on the normal two-particle or four-point vertex contains either two-loop diagrams (for which there is small phase space near the Fermi level) or corresponds to self-energy corrections on internal lines, which are ignored. Third, the frequency dependence of the interaction is ignored and all vertices are studied at zero external frequencies. The wavevector-dependence is treated in the $N$-patch approximation which resolves the angular dependence around the Fermi surface by introducing $N$ sectors with Fermi wave vectors $\vec{k}_i$ labeled by indices $k_i$ with $i=1 \dots N$, but drops the radial dependence. All these simplifications are basically dictated by feasibility, and there is no rigorous statement about their validity. However, there is also no clearly documented failure of these approximations in reasonable context. Therefore the critical scales obtained this way should serve the more qualitative purposes of this paper. 

In this approximation, the flow equations for the coupling function $V^\lambda (\vec{k}_1,\vec{k}_2,\vec{k}_3)$ depending two incoming wave vectors $\vec{k}_{1/2}$ and one outgoing $\vec{k}_3$ read 
\begin{align}
\label{dgl_v}
&\frac{d}{d\lambda}V^\lambda ( \vec{k}_1,\vec{k}_2,\vec{k}_3) =\tau^\lambda_{PP}+\tau^\lambda_{PH,d}+ \tau^\lambda_{PH,cr}\, ,
\end{align}
with the particle-particle channel
\begin{align}
\label{pp}
\tau^\lambda&_{PP}( \vec{k}_1,\vec{k}_2,\vec{k}_3 )\notag\\
=&- T \sum_{\vec{k},ik_0}  V^\lambda ( \vec{k}_1,\vec{k}_2,\vec{k})
\cdot L^\lambda ( k,q_{PP} ) V^\lambda ( \vec{k},\vec{q}_{PP},\vec{k}_3) \, , 
\end{align}
the direct particle-hole channel
\begin{align}
\tau&^\lambda_{PH,d}( \vec{k}_1,\vec{k}_2,\vec{k}_3) \notag\\
= &-T \sum_{\vec{k},ik_0}  \Bigl[-2V^\lambda(\vec{k}_1,\vec{k},\vec{k}_3) L^\lambda(k,{q}_{PH,d} ) V^\lambda(\vec{q}_{PH,d},\vec{k}_2,\vec{k}) \notag\\
&+ V^\lambda(\vec{k},\vec{k}_1,\vec{k}_3) L^\lambda(k,{q}_{PH,d} ) V^\lambda(\vec{q}_{PH,d},\vec{k}_2,\vec{k}) \notag\\
&+ V^\lambda(\vec{k}_1,\vec{k},\vec{k}_3) L^\lambda(k,{q}_{PH,d} ) V^\lambda(\vec{k}_2,\vec{q}_{PH,d},\vec{k})    \Bigr] \, , \label{phd} 
\end{align}
and the crossed particle-hole channel
\begin{align}
\label{phcr}
\tau^\lambda&_{PH,cr}( \vec{k}_1,\vec{k}_2,\vec{k}_3)\notag\\
= &- T \sum_{p,ik_0}  V^\lambda(\vec{k},\vec{k}_2,\vec{k}_3)  L^\lambda(k,{q}_{PH,cr} ) V^\lambda(\vec{k}_1,\vec{q}_{PH,cr},\vec{k}) \, , 
\end{align}
where $k=(ik_0,\vec{k})$ are wave vector and Matsubara frequency of the first line and $\vec{q}_{PP}=-\vec{k}+\vec{k}_1+\vec{k}_2$, $\vec{q}_{PH,d}=\vec{k}+\vec{k}_1-\vec{k}_3$, $\vec{q}_{PH,cr}=\vec{k}+\vec{k}_2-\vec{k}_3$ are the wavevectors of the second loop line. $T$ is the temperature. All incoming and outgoing frequencies are set to zero and the frequency of the second line is fixed by frequency conservation to be $-ik_0$ in the $PP$ diagram and $ik_0$ in the $PH$ diagrams. Also the fourth momentum in the interaction vertex is fixed by conservation, and the spin convention for the coupling function $V^\lambda (\vec{k}_1,\vec{k}_2,\vec{k}_3)$ is that the first incoming line $\vec{k}_1$ and the first outgoing line $\vec{k}_3$ have the same spin projection $\sigma$, while the second incoming $\vec{k}_2$ and the second outgoing have spin $\sigma'$.
The internal loop is given by
\begin{equation}
L^\lambda(k,k')= \frac{d}{d\lambda} \left[ G^\lambda(k) G^\lambda(k') \right] \, ,
\end{equation}
where in  our  approximation  self-energy corrections are neglected, i.e., the full propagator is identical to the free propagator. The fRG cutoff is introduced by multiplying the Green's function by a function that cuts out the low-energy modes,
\begin{equation}
G^\lambda(k)  = \frac{C^\lambda [E(\vec{k})] }{ik_0 - E(\vec{k}) } \; . 
\end{equation}
In the numerical implementation of the flow scheme, we use a slightly smoothened step function, $C^\lambda [E(\vec{k})]  \approx \Theta \left( |E(\vec{k})|  - \lambda \right) $. 
The initial condition for the flow at the scale $\lambda_0$ of the band width of the conduction band is given by the orbital-content dressed coupling function in Eq. (\ref{omakeup}), i.e., by
\begin{equation}
V^{\lambda_0}(\vec{k}_1,\vec{k}_2,\vec{k}_3) = V_{cccc} (\vec{k}_1,\vec{k}_2,\vec{k}_3) \, ,
\end{equation}
with band index $c$ for the $d_{x^2-y^2}$-dominated conduction band.

The integration of the above fRG equation takes into account all one-loop corrections with both internal lines in the low-energy window and thereby reconstructs the most important parts from the perturbation expansion of the band near the Fermi level. Usually, the fRG flow leads to strong coupling, i.e., at some low critical energy scale $\lambda_c$, at least  one family of couplings $V^\lambda (\vec{k}_1,\vec{k}_2,\vec{k}_3)$ grows very large, and the perturbative flow without self-energy feedback has to be stopped. The different types of divergences for the single-band Hubbard model as function of the system parameters have been analyzed and classified in a number of previous publications (see, e.g., Refs. 17 and 18). In the present case the fRG flow almost exclusively favors a $d$-wave pairing instability that is understood to be driven by antiferromagnetic spin fluctuations. In this paper we do not discuss any other features of this $d$-wave instability except its critical scale $\lambda_c$. This scale can be understood as a measure or upper estimate for the transition temperature $T_c$ into the $d$-wave superconducting state. It takes an analogous  role as the quantity $\hbar \omega_D e^{1/\rho_0 g}$ in a simple BCS problem with Debye frequency $\omega_D$, density of states $\rho_o$ and attractive interaction $g$.

So far, the bands away from the Fermi surface have only entered through the orbital content. Let us next discuss the first step to include virtual processes into these bands, i.e. inter-band transition. If we write a perturbation series for the full two- or three-orbital model, and compare the various one-loop terms entering there to what we get in the fRG for the conduction band, we find that the next important terms that are not included in the fRG treatment are diagrams with at least one internal line in one of the bands away from the Fermi surface. While these diagrams will not become singular as their energy denominator will never go to zero, they might induce additional wave vector dependencies to the interactions, that might be measurable for the unconventional pairing studied here. Actually, the constrained-RPA formalism sums up a part of these contributions to infinite order, which results in an additional screening of the Coulomb interaction.\cite{crpa,crpa2} Including these effects more systematically into the fRG for a single band is a separate issue that we do not touch here. Instead, we will study the impact of an additional second-order correction
 due to these virtual processes with one and also two internal lines away from the conduction band, whereas the contribution of the latter is confirmed to be comparably small. This correction, which we will refer to as {\em high-energy second-order correction}, is added to the initial condition of the flow in the conduction band. It is given by all one-loop diagrams (particle-particle and particle-hole) with bare interactions at the vertices and at least one internal line in a band above or below the conduction band. We will see that it has a definite impact for smaller energy gaps between the bands.
Due to the energetic separation of the bands away from the Fermi surface, we expect that this second-order treatment is already quite good, and that the higher-order corrections neglected here only have a small effect.

\section{Results for the two-band model}
\label{twoband}
Let us now apply the $N$-patch fRG scheme to the lower band of the two-band model described in Sec. II. We therefore fix the hopping parameters of the bands to be $t_d=0.45 eV$, $t_s=0.5 eV$, and the hybridization to  $t_{sd}=0.6 eV$. The tuning parameter is the energy separation between the band centers, $\Delta E_{sd}$. 

\begin{figure}
\begin{center}
\includegraphics[scale=0.25]{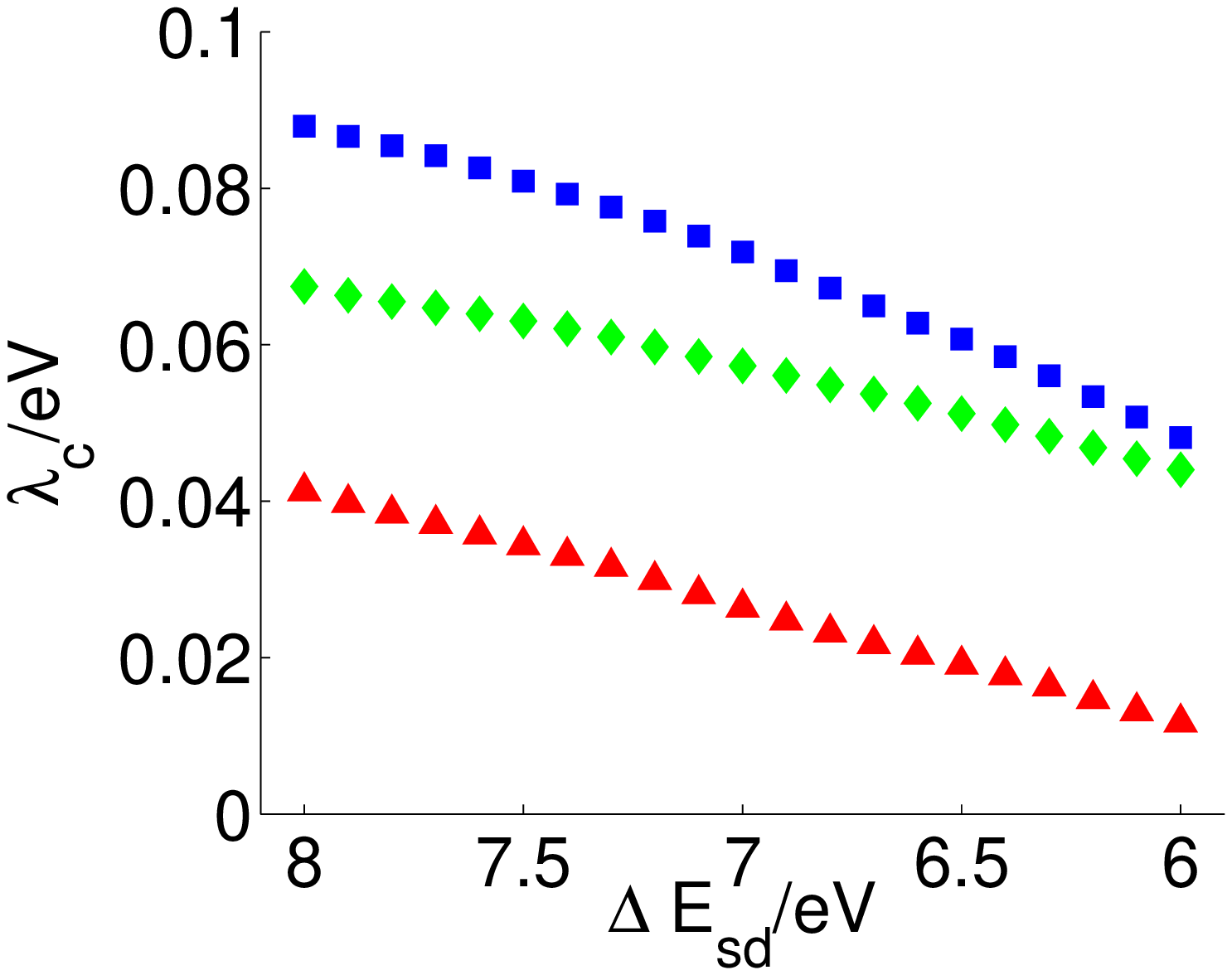}
\includegraphics[scale=0.25]{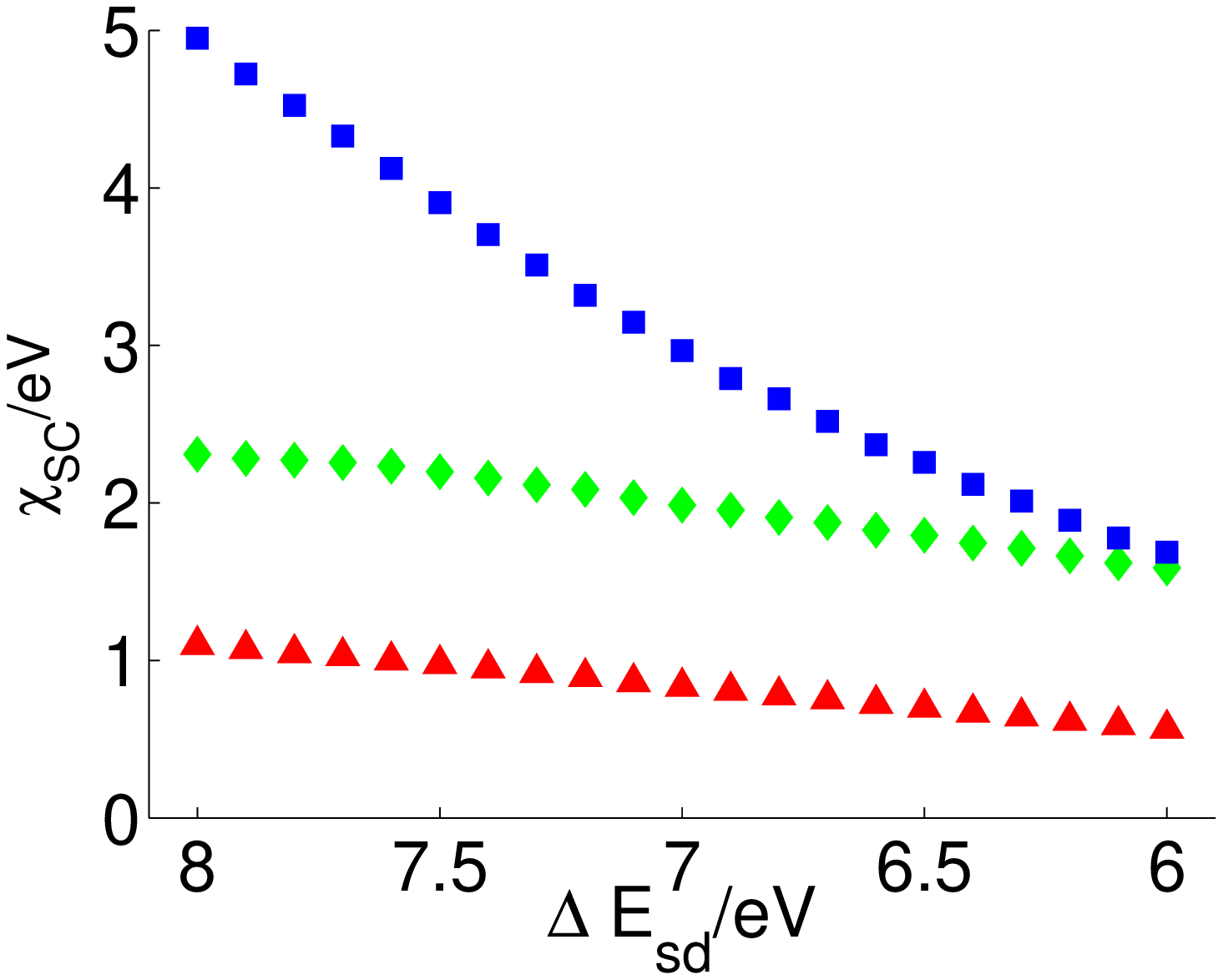}
\end{center}
\caption{Critical scale (left) and superconducting susceptibility at a fixed scale $0.09 eV $ (right) vs $\Delta E_{sd}$, computed with 64 patches on the Fermi surface. Red triangles: with orbital mixing, green diamonds: orbital mixing cut out by hand, $U=1.8 eV$, $U'=1.44 eV$, $J_H=J_P=0.18eV$. Blue squares: with larger interactions $U=2.2 eV$, $U'=1.76 eV$, $J_H=J_P=0.22eV$, with orbital mixing. Parameters: $t_d=0.45 eV$, $t_s=0.5 eV$, $t_{sd}=0.6eV$, $\langle n \rangle=0.84$.}
\label{2bandplot} 
\end{figure}

In the plots of Fig. \ref{2bandplot} we additionally choose two sets of interaction parameters according to the sum rule $U-U'=2 J_H=2J_P$. \cite{dagotto} The density is fixed to 16$\%$ hole doping of the lower band, i.e. $\langle n \rangle =0.84$ per site. In the left plot, the critical scale for the $d$-wave pairing instability, $\lambda_c$,  in dependence of the energy difference $\Delta E_{sd}$ of the two orbitals is shown. The direction of the horizontal axis is inverted so that the plots can be compared better with the ones for the three-band model in Sec. \ref{sectthreeband}. We can see that moving the two bands closer to each other, i.e., reducing $\Delta E_{sd}$, decreases the critical scale $\lambda_c$. This behavior is robust with respect to the interaction strength. We also plot $\lambda_c$ vs $\Delta E_{sd}$ when the orbital content is ignored and the bare interaction is simply $U$ for all wavevector combinations. In this case, $\lambda_c$ is higher. This allows us to distinguish the effect of the orbital content from that of the Fermi surface shape, which also changes as a function of the energy separation $\Delta E_{sd}$. Here, we only calculated data for the parameter set with smaller interaction, as the one-loop approximation is not justified, if the interaction is to large. With included orbital mixing however the average of the coupling function at the beginning of the flow is below half of the bandwidth. Summarizing our findings, we can state that the expected decrease of $\lambda_c$ with the Fermi surface becoming rounder when $\Delta E_{sd}$ becomes smaller is not compensated by the orbital content or stronger hybridization in the $d_{x^2-y^2}$-band. Rather, this latter effect leads to an additional reduction of the pairing scale. 

We also plot in Fig. \ref{2bandplot} the effective strength of the $d$-wave superconducting channel at a small fixed $\lambda > \lambda_c (\Delta E_{sd})$, which is defined as
\begin{equation}
\label{sussl}
\chi_{\text{SC}}=\frac{1}{N^2}\sum_{\vec{k},\vec{p}}
V^\lambda ( \vec{k},-\vec{k},\vec{p}  ) f(\vec{k}) f(\vec{p}) \, .
\end{equation}  
Here, the sums run over all $N$ patches of our patching scheme and $f(\vec{k})$ and $f(\vec{p})$ are $d$-wave form factors and are given by $f(\vec{k})=\cos(k_{x})-\cos(k_{y})$, and  $V_\lambda \left( \vec{k},-\vec{k},\vec{p}  \right) $ is the running interaction at the small fixed scale. This measure of the strength of the superconducting channel gives a qualitatively similar picture. Again, both factors, Fermi surface deformation and orbital makeup, reduce the tendency toward $d$-wave pairing.

In this sense, a well separated $d_{x^2-y^2}$-like band appears to be the optimal situation for a high critical scale or high $T_c$.
The two-band model with the $s$ orbital above the $d_{x^2-y^2}$ orbital can not give an explanation why the cuprates with rounder Fermi surface have higher $T_c$, as indicated by the DFT trends. Our data clearly show that the perturbation of the dispersion as well as the orbital content reduce the critical scale.

In order to understand this finding more directly, and to see how general it is, we will now turn to a simplified description in terms of the two-patch model.

\section{Simplified picture: Two-patch model}
\label{sect-twopatch}
Let us consider again the situation where only the lower $d_{x^2-y^2}$-like band has a Fermi surface, and let this Fermi surface be near the van Hove points at $A=(\pi,0)$ and $B=(0,\pi)$.  Then a common approximation is the so-called two-patch model,\cite{twopatch} where only the fermionic degrees of freedom in small circles around these van Hove points are kept, and the interactions between these regions are approximated by four constants $g_1$ to $g_4$. These are defined by
\begin{align}
g_1 & =V(A,B,B,A)=V(B,A,A,B) \, ,\\
g_2 & =V(A,B,A,B)=V(B,A,B,A) \, ,\\
g_3 & =V(A,A,B,B)=V(B,B,A,A) \, ,\\
g_4 & =V(A,A,A,A)=V(B,B,B,B) \, ,
\end{align}
where the notation, is again, that the first two entries $\vec{k}_1$ and $\vec{k}_2$ of $V(\vec{k}_1,\vec{k}_2,\vec{k}_3,\vec{k}_4)$ are the incoming wave vectors, and that the first incoming particle with $\vec{k}_1$ and the first outgoing with $\vec{k}_3$ have the same spin projection $\sigma$, while the second incoming $\vec{k}_2$ and the second outgoing $\vec{k}_4$ have $\sigma'$.
Such a modeling has been used to explore the basic phase diagram of the two-dimensional single-band $t$-$t'$ Hubbard model near van Hove filling. There, the initial value for the $g_i$ from the bare interaction is just $U$.  Recently, a very similar model has been used for the study of the iron arsenide superconductors.\cite{chubukov}

The fRG treatment of the two-patch model for the Hubbard model, where only the singular one-loop contributions of the particle-particle loop with zero incoming momentum and the particle-hole loop with momentum transfer $(\pi,\pi)$ are kept, leads to the flow equations 
\begin{eqnarray}
\dot{g}_1&=&2 \dot{d}_1 g_1(g_2-g_1) \label{g1flow} \, ,\\
\dot{g}_2&= &\dot{d}_1 (g_2^2+g_3^2) \label{g2flow} \, ,\\
\dot{g}_3&=&-2 \dot{d}_0 g_3 g_4 +2 \dot{d}_1 g_3 (2 g_2-g_1) \label{g3flow} \, ,\\
\dot{g}_4&=&-\dot{d}_0(g_3^2 +g_4^2) \, , \label{g4flow}
\end{eqnarray}
with $\dot{d}_0,\dot{d}_1>0$ as scale derivatives of the particle-particle and particle-hole loop, respectively.
Possible instabilities in the $d$-wave pairing, antiferromagnetic spin-density wave (SDW), and charge-density wave channel are then associated with the unbounded growth of the combinations $g_3-g_4$, $g_2+g_3$ and $-2g_1+g_2-g_3$, respectively. In the one-band Hubbard model for $d_0=d_1$, corresponding to perfect nesting,  the coupling constants diverge as $g_{2,3} \rightarrow \infty$, $g_4 \rightarrow -\infty$, $g_1$ diverges more slowly. 
As long as $g_4$ is positive, the two terms on the RHS of Eq. (\ref{g3flow}) have opposite sign and give competing contributions, but eventually $g_4$ becomes negative and both terms in Eq. (\ref{g3flow}) drive the flow to strong coupling. Then the $d$-wave pairing $g_3-g_4$ diverges, but usually, for nonzero $U$, the SDW combination $g_2+g_3$ has grown much larger. Hence the instability for $d_1=d_0$ is of multichannel type, with the SDW tendencies being strongest.
For $d_1<d_0$, corresponding to  nonzero $t'$, the SDW channel is less dominating and one gets a regime where the $d$-wave coupling $g_3-g_4$ diverges more strongly. This is what we call the $d$-wave pairing regime. A more detailed analysis is given in Ref. \onlinecite{twopatch}.

For the single-band model, the comparison with the fRG calculations taking into account the full wave vector dependence of the interactions around the Fermi surface showed that most trends can already be inferred from the two-patch model, so that the latter should serve as a good starting point to study the main effects.

We will now use the two-patch model to assess the impact of the orbital content on the type and energy scale of the instability of the weakly-coupled state due to the interactions. At the van Hove points, the transformation that diagonalizes the hopping Hamiltonian of the two-band model can be expressed as 
\begin{align}
u_{cd}(A) =\cos \phi , \quad & u_{cd}(B)=\cos \phi , \\
u_{cs}(A)=\sin \phi , \quad & u_{cs}(B)=-\sin \phi ,  
\end{align}
where $u_{cs}$ and $u_{cd}$ are the orbital weights of $s$ and $d_{x^2-y^2}$ orbitals in the conduction band $c$, which is the one with the Fermi surface near the van Hove points. The angle $\phi$ is a measure for the strength of the orbital mixing and can be understood as the angle between the vertical and the arrow at the van Hove points in Fig. \ref{2bandplots}. In the case of $\phi=0$, the mixing is zero and we recover the pure one band model without mixing effects. The sign change of the $s$ admixture in $u_{cs}(B)$ compared to $u_{cs}(A)$ comes from the different in-plane symmetry of the $d_{x^2-y^2}$ orbital. It is crucial for the following. Note that this sign change would also occur for any other orbital that is admixed to the $d_{x^2-y^2}$ orbital that transforms trivially under 90 degree rotation in the plane, i.e. also for $d_{z^2}$ or $p_z$ orbitals. Hence, qualitatively, the results hold more generally for a wider class of admixed orbitals.

 Next we consider here only local intra- and interorbital repulsion $U$ and $U'\le U$ to keep it simple. Hund's couplings and pair hopping terms in a realistic range are smaller than these parameters, and should not lead to qualitative differences in the results. 
 For the intraorbital interaction, the orbital content gives a factor $\cos^4 \phi$, while for the interorbital interaction $U'$ the two incoming and the two outgoing particles are in different orbitals, contributing $\pm 2 \sin^2 \phi \cos^2 \phi$. The sign now depends on whether the particle in the $s$ orbital remains in the same patch, say $A$, or whether it gets scattered to the other patch, say from $A$ to $B$.
In total, we obtain for the initial values of the four coupling constants:
\begin{align}
g_{1,3} & =U \, \left( \cos^4 \phi+\sin^4 \phi \right) - 2 U' \cos^2 \phi \sin^2 \phi  \, , \\
g_{2,4} & =U \, \left( \cos^4 \phi+ \sin^4 \phi�\right) + 2 U'   \cos^2 \phi \sin^2 \phi \, .
\end{align}
We see that the orbital content $\phi \not= 0$ in this $d_{x^2-y^2}$-case quite generally suppresses the bare interaction strength. For a nonzero value of $U'>0$, this suppression is stronger for $g_{1,3}$ and weaker for $g_{2,4}$. Generally, a larger $g_2$ and $g_3$ would lead to a higher critical scale for $d$-wave pairing, as these couplings drive the antiferromagnetic fluctuations that form the pairing glue.  More repulsive $g_1$ and $g_4$ leads to a smaller critical scale. For the desired $d$-wave pairing instability, the relative enhancement of $g_4$ with respect to $g_3$ for $\phi \not=0$ is exactly the wrong way. It corresponds to an additional repulsion in the $d$-wave pairing channel and reduces the critical scale.

\begin{figure}
\begin{center}
\includegraphics[scale=0.3]{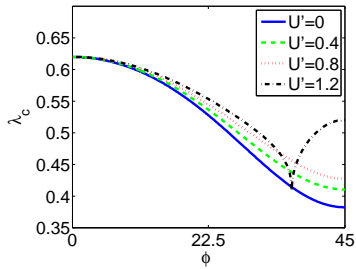}
\end{center}
\caption{Critical scale of the two-patch model as a function of the mixing parameter $\phi$ for $U=1$ and different $U'$. 
For most parameters, the instability is of antiferromagnetic spin-density wave or $d$-wave pairing type, and the critical scale decreases with larger admixture $\phi$. Only for $U'>U$ and $\phi$ large enough, the instability turns into a charge-density wave instability, and the critical scale is increased by further increasing $\phi$ (black dashed dotted curve).}
\label{gusplot} 
\end{figure}

We have performed a numerical analysis of the flow equations (\ref{g1flow})--(\ref{g4flow}), in order to investigate the influence of the mixing on the critical scale. 
Fig. \ref{gusplot} shows the critical scale obtained for the two-patch model for different $U'$. It turns out that the critical scale is suppressed by mixing as long as $U'$ does not get too large. Only for $U'>U$, we encounter the charge-density wave instability and here the $s$ admixture $\phi \to \pi/2$ can actually increase the critical scale in a certain range. Here, the initial value of $g_3$ becomes negative and thus $g_3$ flows to $-\infty$, as the second term of the right hand side of Eq. (\ref{g3flow}) is negative and the first term eventually becomes negative when $g_4<0$. 
In this numerical treatment, we assumed that the relevant loops $d_0$ and $d_1$ have the same value during the flow, which should be an acceptable assumption for these qualitative conclusions. However we checked that the qualitative results do not depend on the individual behavior of the loops during the flow. 

Thus we conclude that already the simple two-patch model can give us a simple explanation how orbital mixing effects lower the critical temperature of a $d$-wave superconducting instability. 

\section{Results for the Three-band Model}
\label{sectthreeband}
So far, we have seen that admixing a single orbital to the $d_{x^2-y^2}$-derived band leads to a reduction of the critical scale for $d$-wave pairing. This causes problems to understand the mentioned apparent material trend in the high-$T_c$ cuprates that lowering an axial orbital in energy from above results in higher critical temperatures. Here, we will show that this trend can be recovered if additional orbitals are considered. 
To this end, we now consider a two-dimensional three-band model, which includes the $d_{x^2-y^2}$, $d_{z^2}$, and $s$ orbitals. The Hamiltonian of this model is given in Eq. (\ref{3band-hamiltonian}). 

To find out whether the mentioned material trend can be explained by mixing effects, we study the critical scale for $d$-wave pairing with different values for the onsite kinetic energy of the $d_{z^2}$ and $s$ orbitals. 
We follow the path through the parameter space proposed by Sakakibara \textit{et al.}\cite{sakakibara} More precisely, we increase  $\Delta E=t^d_{0,0}-t^z_{0,0}$ by hand but leaving the difference between $d_{z^2}$ and $s$ orbital constant.  This is motivated by the observation that in the cuprate HgBa$_2$CuO$_4$ with $T_c \sim 90 \text{K}$ both the $d_{z^2}$ and $s$ orbital are lowered with respect to the $d_{x^2-y^2}$ orbital compared to La$_2$CuO$_4$ with lower $T_c \sim 30 \text{K}$. HgBa$_2$CuO$_4$ has a higher $T_c$ and a more rounded Fermi surface than La$_2$CuO$_4$. As the change in $\Delta E$ appears to be the most striking effect, we do not consider a variation of other hopping parameters to keep our analysis simple. Note that a larger $\Delta E=t^d_{0,0}-t^z_{0,0}$ thus corresponds to a smaller $\Delta E_{sd}$ of Sec. \ref{twoband}, which is written as $\Delta E_{sd}=t^s_{0,0}-t^d_{0,0}$ in terms of the three-band model in Eqs. (\ref{3band-hamiltonian})--(\ref{hoppings2}). All calculations presented in this section are done at $16\%$ hole doping, that is, $\langle n \rangle=2.84$ per site.

\begin{figure}
\begin{center}
\includegraphics[scale=0.25]{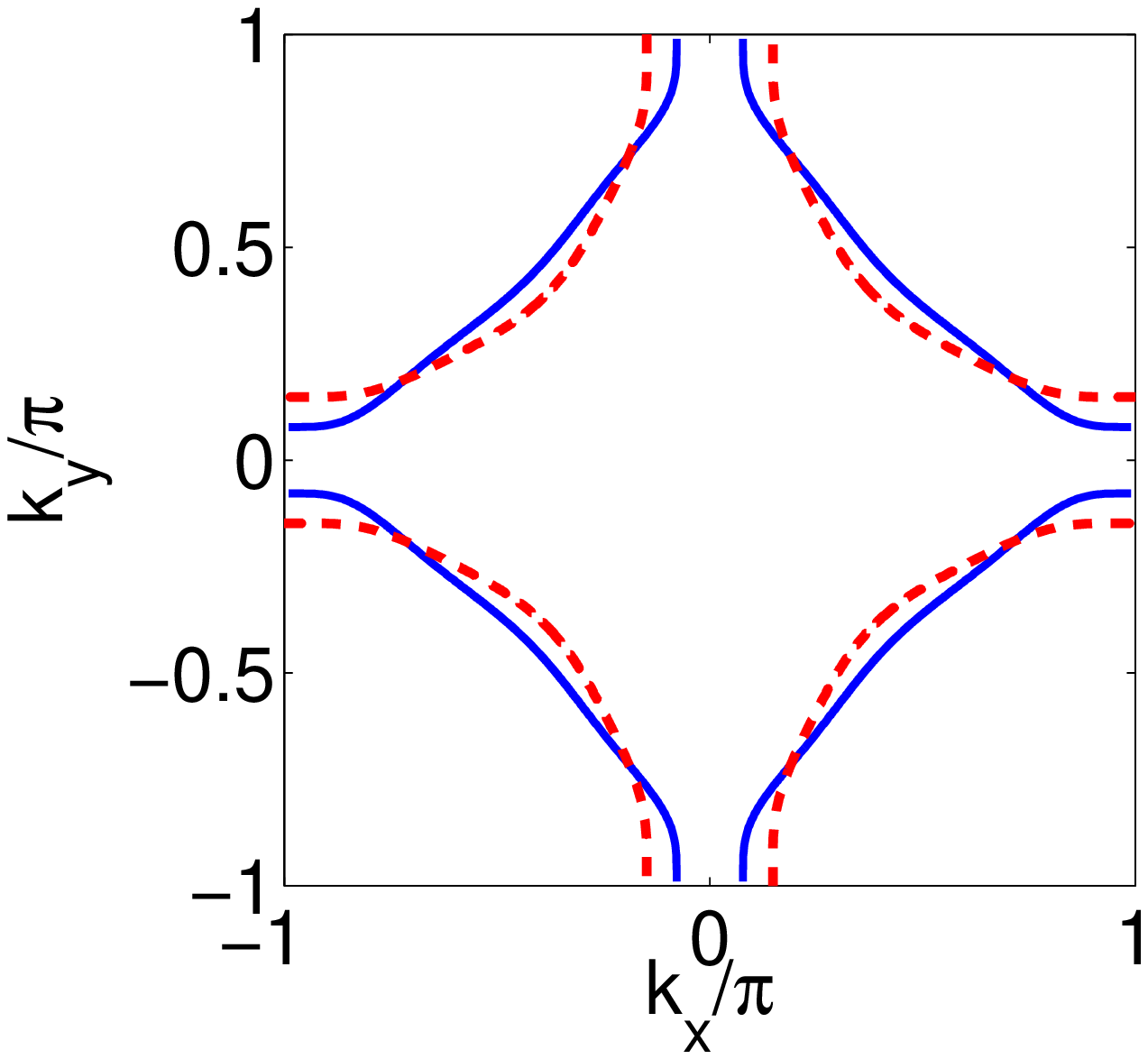}
\includegraphics[scale=0.25]{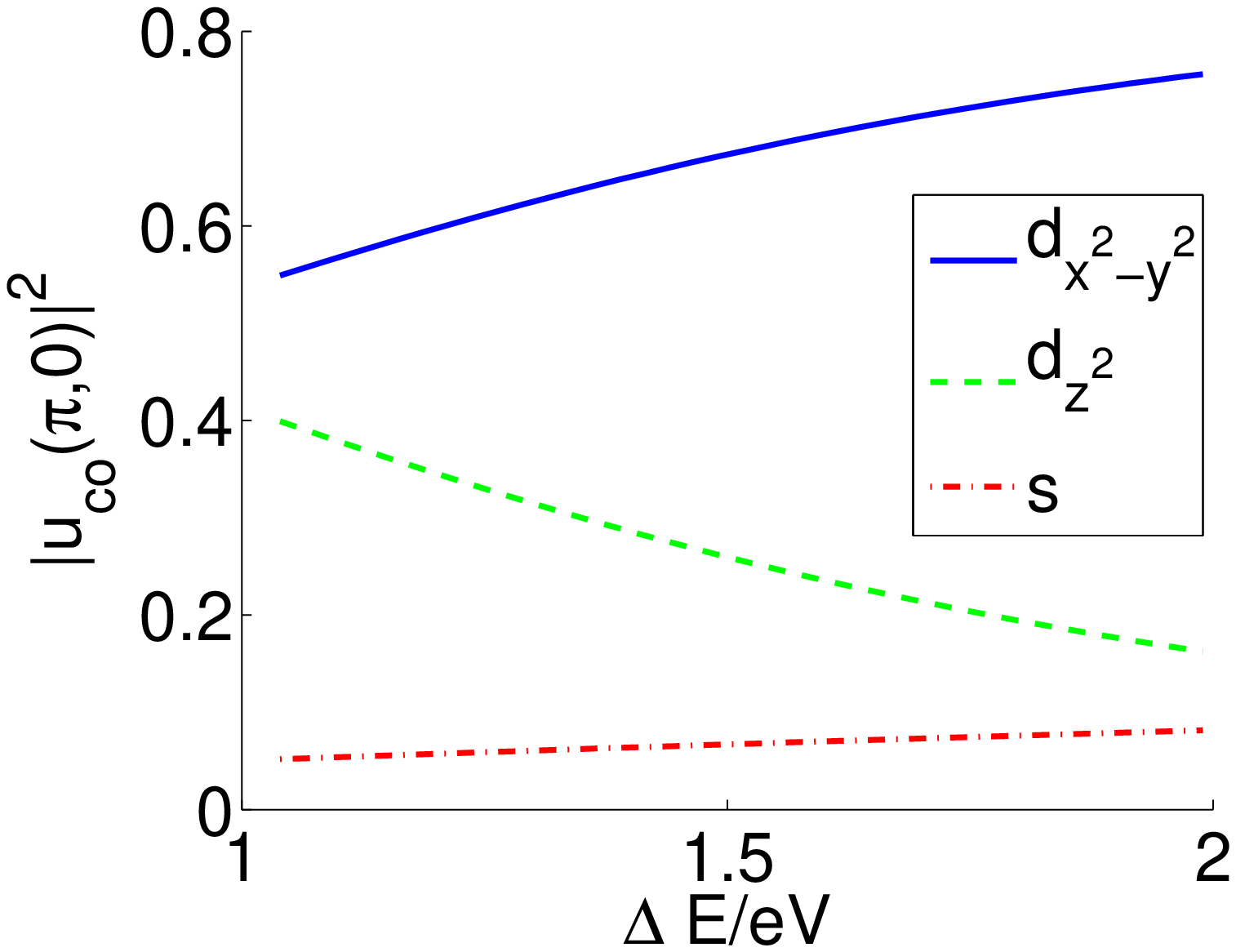}
\end{center}
\caption{Left: shape of the Fermi surface for parameters corresponding to La$_2$CuO$_4$ (blue solid line) and for $\Delta E\sim 2eV$ (red dashed line), corresponding to HgBa$_2$CuO$_4$, $\langle n \rangle=2.84$. Right: admixture of the three orbitals to the $d_{x^2-y^2}$-dominated band with the Fermi surface, the orbitals $d_{x^2-y^2}$, $d_{z^2}$ and $s$ are shown as a blue solid, green dashed, and red dotted lines, respectively.}
\label{fsplot} 
\end{figure}

The left plot of Fig. \ref{fsplot} shows the Fermi surface of our model for the two situations. While the $s$ orbital lowers the energy of the van Hove point, the $d_{z^2}$ orbital has the opposite effect because it lies below the $d_{x^2-y^2}$ orbital. Thus the combined effect of the $s$ and $d_{z^2}$ orbitals can lead to a roughly squarelike Fermi surface, as it would be without mixing at all. The plot on the right shows the relative orbital admixture at the van Hove point for different $\Delta E$. $|u_{co}(\pi,0)|^2$ is the squared absolute value of the coefficients in the orbital-band transformation given in Eq. (\ref{eqtrafo}), where the index $c$ as above denotes the $d_{x^2-y^2}$-dominated conduction band. This quantity serves as a measure for the orbital admixture. The contributions from the three orbitals add up to one due to the unitarity of the transformation. 
With larger $\Delta E$ the Fermi surface becomes more rounded, due to a larger perturbation from the $s$ orbital and a smaller influence from the $d_{z^2}$ orbital. The total mixing, i.e., the non-$d_{x^2-y^2}$-content, is weakened with larger $\Delta E$. This mainly comes from the decreasing influence of the $d_{z^2}$ orbital when it moves down in energy. On the other hand, the $s$ orbital gets closer to the $d_{x^2-y^2}$ orbital with larger $\Delta E$ and thus increases the mixing. This effect, however, is much smaller due to the larger gap between $d_{x^2-y^2}$- and $s$ orbitals, and because the $s$ band is rather wide. We conclude that the lowering of $\Delta E$ leads to two competing effects on the critical scale. First, the Fermi surface gets more rounded. which according to the results in one-band models decreases the critical scale. Second, and this is now different from the two-orbital case, the orbital mixing is reduced, and the $d_{x^2-y^2}$ character increases. As we have seen in the previous section, this favors an enhancement of the critical scale.

\begin{figure}
\begin{center}
\includegraphics[scale=0.25]{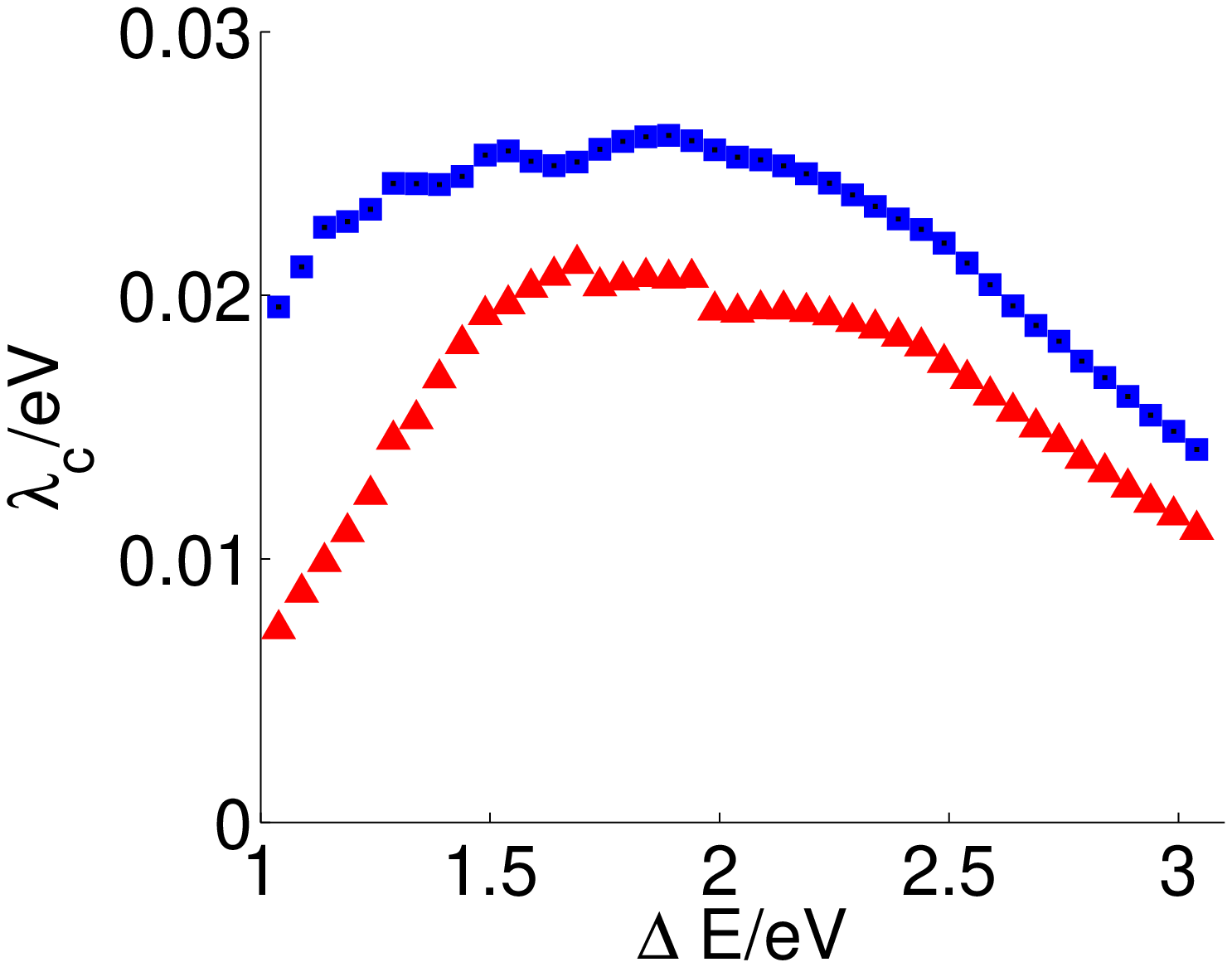}
\includegraphics[scale=0.25]{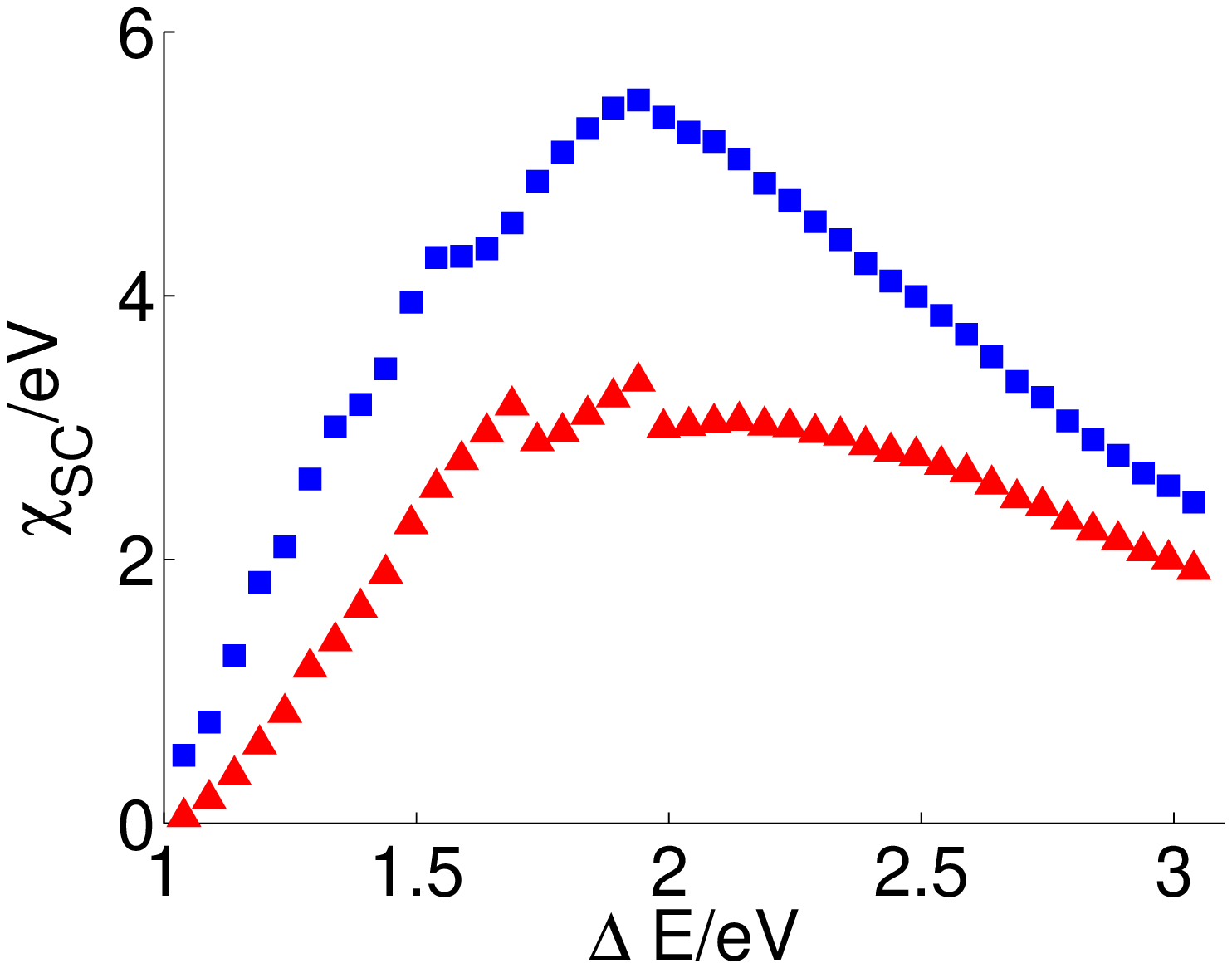}
\end{center}
\caption{fRG data for the three-band model with interaction parameters $U=2.2eV$, $U'=1.76eV$, $J_H=J_P=0.22eV$ computed with 64 patches, $\langle n \rangle=2.84$. Left plot: critical scale $\lambda_c$ vs $\Delta E=t^d_{0,0}-t^z_{0,0}$. Right plot: average $d$-wave pairing at a fixed scale $\lambda=0.027eV$ vs $\Delta E$; without high-energy second-order correction (blue squares), with high-energy second-order correction (red triangles), explained in Sec. III.}
\label{3bandplot1} 
\end{figure}
\begin{figure}
\begin{center}
\includegraphics[scale=0.25]{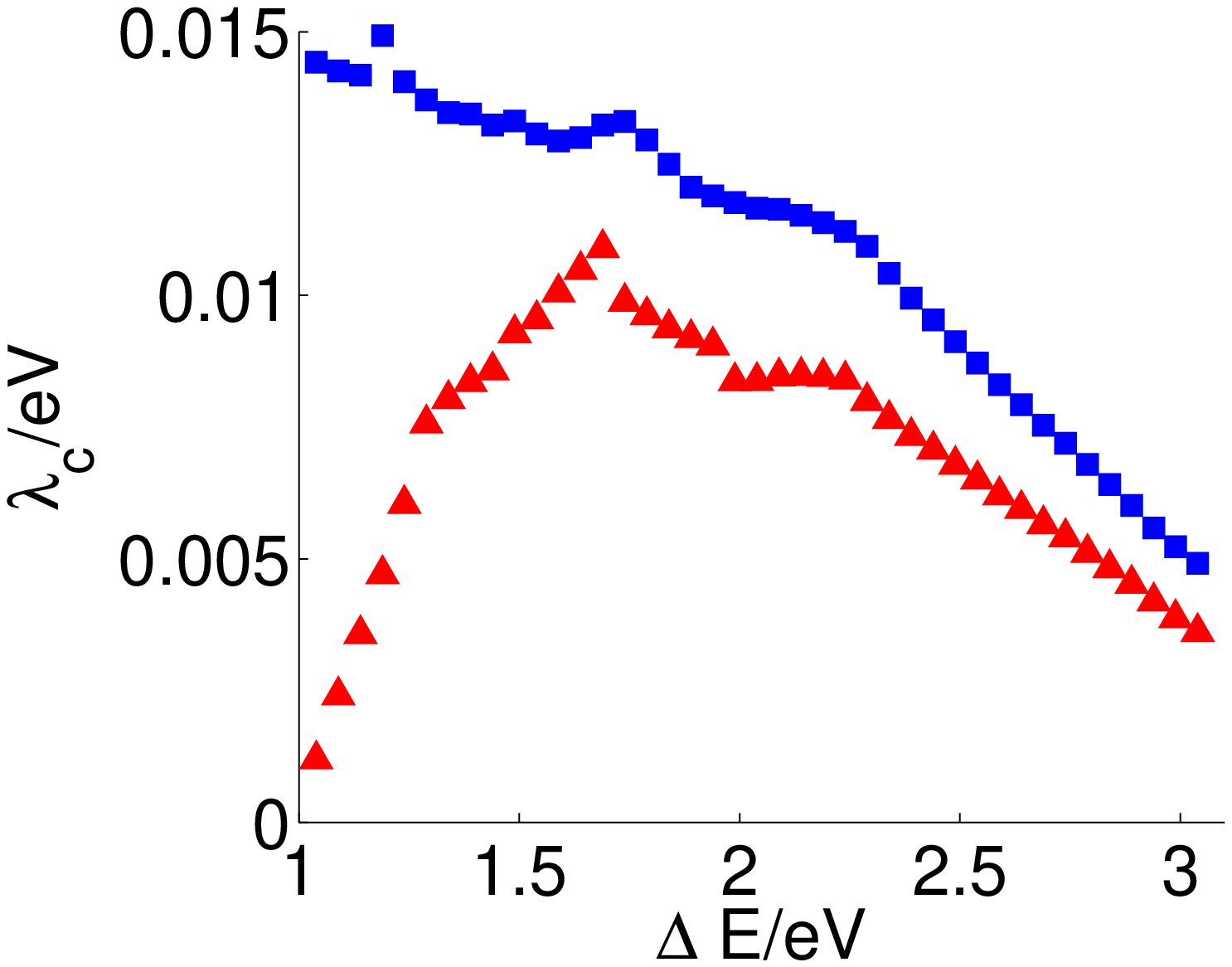}
\includegraphics[scale=0.25]{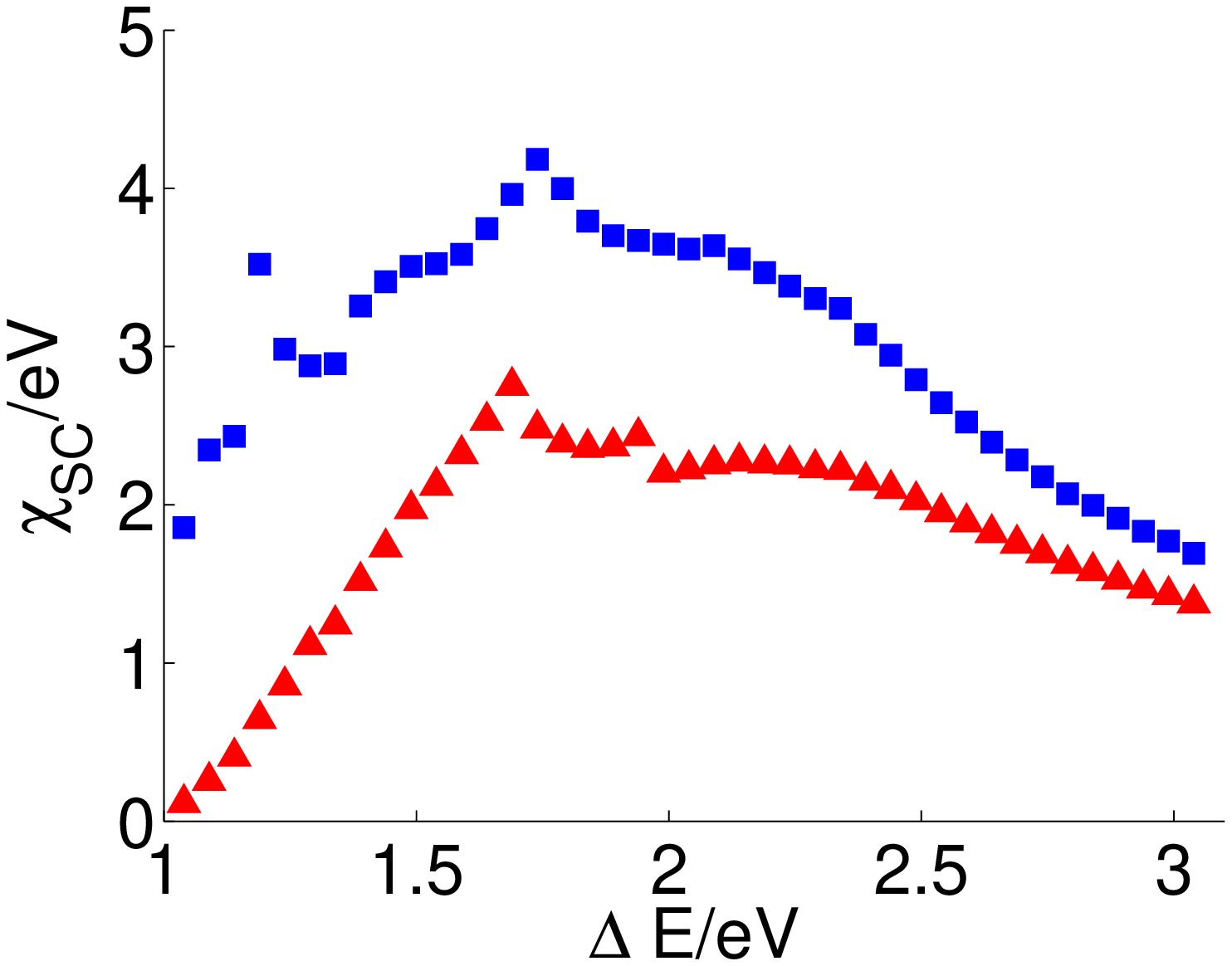}
\end{center}
\caption{fRG data for the three-band model with interaction parameters $U=2.0eV$, $U'=1.6eV$, $J_H=J_P=0.2eV$ computed with 64 patches, $\langle n \rangle=2.84$. Left plot: critical scale $\lambda_c$ vs $\Delta E=t^d_{0,0}-t^z_{0,0}$. Right plot: average $d$-wave pairing at a fixed scale $\lambda=0.015eV$ vs $\Delta E$; without high-energy second-order correction (blue squares), with high-energy second-order correction (red triangles), explained in Sec. III. The steps in the data are due to the discretization of the wave vector dependence of the coupling function.}
\label{3bandplot2} 
\end{figure}

In Fig. \ref{3bandplot1}, the critical scale, obtained from the fRG calculation, is shown. Here, the interaction parameters are chosen to be $U=2.2eV$, $J_H=U_{P}=0.1 U$, and $U'=0.8 U$. We use 64 patches to discretize the Brillouin zone according to the patching scheme described above.\cite{hsfr} With the interaction parameters chosen here the average coupling constant at the beginning of the flow is below half of the bandwidth. Both approximations, with and without high-energy second order correction, show the same qualitative trend.
The two competing effects now induce a maximum in the critical scale and $\chi_{\text{SC}}$ at some $\Delta E^*$. For larger $\Delta E> \Delta E^*$ the reduced orbital mixing of the vertices cannot compensate the change of the Fermi surface. To separate the effect of the orbital mixing on the interaction and the dispersion, we run the same fRG flow, but now starting  with a simple onsite repulsion regardless of the orbital mixing. In other words, we cut out the mixing effect on the vertices by hand. As expected from the results of the two-patch model the critical scale without the orbital content is indeed drastically increased and we do not get an enhancement of the critical scale with increasing $\Delta E$, because now only the Fermi surface effect of the shift comes into play. 

Additionally we run the fRG flow again but now shift only the  $s$ or the $d_{z^2}$ orbital respectively to see, if the interpretation suggested from the two-patch-model holds. As expected, if we only lower the $s$ orbital, the critical scale decreases, since the mixing of the $d_{z^2}$ orbital is not reduced and the positive effect on the critical scale is not present. If we on the other hand lower only the $d_{z^2}$ orbital but keep the $s$ orbital constant, the qualitative picture from Fig. \ref{3bandplot1} is recovered.

The value $\Delta E^*$ with highest critical scale depends on the strength of the interactions. For smaller bare values, the Fermi surface shape is more important, and $\Delta E^*$ shifts to smaller values and even disappears. In Fig. \ref{3bandplot2}, we again show critical scale and $d$-wave pairing strength vs $\Delta E$ in Fig. \ref{3bandplot1}, but now for reduced onsite interaction $U=2eV$, while the other interaction parameters are reduced accordingly so that the ratio to the onsite interaction remains constant. Now without second-order correction of the additional orbitals the critical scale decreases over the entire range. Our data quite generally suggests that at least without high-energy corrections the effect of an enhancement of the critical scale with larger $\Delta E$ is less pronounced, when the critical scale is lower, that is for smaller interaction strength or, e.g., larger $J_H$, which usually suppresses the flow to strong coupling. Turning this observation around, we can argue that the enhancement effect might even be larger if we used realistic interaction values for the cuprates. We refrain from running the fRG flow with such large interaction parameters, as the one-loop approximation of the fRG equation is then no longer justified. 

It can be seen that the inclusion of the high-energy vertices within second-order perturbation can have a profound effect on the critical scale. Especially for smaller interaction strength, the instability is strongly suppressed at small $\Delta E$, so that in contrast to the case without high-energy correction even at smaller interaction strength a maximum in the critical scale can be found. It seems worthwhile to study the renormalization of the vertices due to higher energy modes in more detail. This may be an additional contribution to the observed material trends, as the curves with included high-energy corrections exhibit a stronger increase in the critical scale than the ones without.

We can compare our findings and our idea, why this $T_c$-trend occurs, with the FLEX two-orbital and three-orbital calculations by Sakakibara \textit{et al.}\cite{sakakibara} Although their Eliashberg eigenvalues cannot be compared directly to the fRG critical scales, it seems that we can qualitatively reproduce their results. Clearly, at least for their two-band calculation, the orbital admixture to the conduction band is certainly lower for the higher $T_c$'s with the rounder surfaces, supporting our explanation. Quantitatively, our results are sensible to the interaction strength and a direct comparison would require data at larger onsite interaction of $3eV$ as used for the FLEX calculation. As the band width of the conduction band is only $\sim 4eV$ we refrain from running the fRG for such large interactions. Also we expect that self-energy effects, which were included by Sakakibara \textit{et al.} but not in our work, will probably still cause a noticeable quantitative difference. 

Within this downfolded three-band electronic structure, HgBa$_2$CuO$_4$ with $T_c$ of $90 \text{K}$ would roughly correspond to $\Delta E =2eV$, close to where the maximal critical scale occurs in Fig. \ref{3bandplot1}. While this is promising, we note that the band structure change from La$_2$CuO$_4$ to HgBa$_2$CuO$_4$ may be more complex, and taking into account these additional changes may affect the observed enhancement as well.  In any case, on a quantitative level and possibly different from Ref. \onlinecite{sakakibara}, our weak-coupling fRG studies cannot be expected to explain the experimental trend, because the real cuprates are more strongly interacting. We propose that strong-coupling methods should be used to find out whether the picture drawn here works the same (and quantitatively better) at strong coupling. Nevertheless, we have identified $\Delta E$ as an important tuning parameter.

The model with the La$_2$CuO$_4$-like band structure parameters has a nested Fermi surface. Consequently, aside from the superconducting instability, we observe a strong tendency toward long-range antiferromagnetic (AFM) order. With larger $\Delta E$, the nesting of the Fermi surface becomes poorer, so that the AFM tendency is more and more suppressed. To have a closer look  on the two competing instabilities, we compare the effective strengths of the respective channels. The $d$-wave superconducting channel can be measured by $\chi_{\text{SC}}$ given in Eq. (\ref{sussl}). Similarly, the AFM channel is 
\begin{equation}
\label{susafm}
\chi_{\text{AFM}}=\frac{1}{N}\sum_{\vec{k}}
V^\lambda \left( \vec{k},\vec{k}_2  \approx (\pi,0), \vec{k}_3 \approx (0,\pi)  \right),
\end{equation}
where  $\vec{k}_2$ and $\vec{k}_3$ are chosen so that they lie close to the two van Hove points and the corresponding momentum transfer is $\vec{k}_2-\vec{k}_3 \approx (\pi,\pi)$. The corresponding ordering susceptibilities diverge if these averaged  strengths of the channels diverge. Thus the defined quantities can be regarded as a measure for the strength of the respective instabilities. The left plots of Figs. \ref{3bandplot1} and \ref{3bandplot2} show that the superconducting channel has, as expected, a similar behavior as the critical scale as function of $\Delta E$. However at $U=2eV$ a qualitative difference arises in the effective strength of the superconducting channel in comparison with the critical scale (see Fig. \ref{3bandplot2}). The effective strength still exhibits a maximum, although it is not as distinct as for larger interactions. 

\begin{figure}
\begin{center}
\includegraphics[scale=0.23]{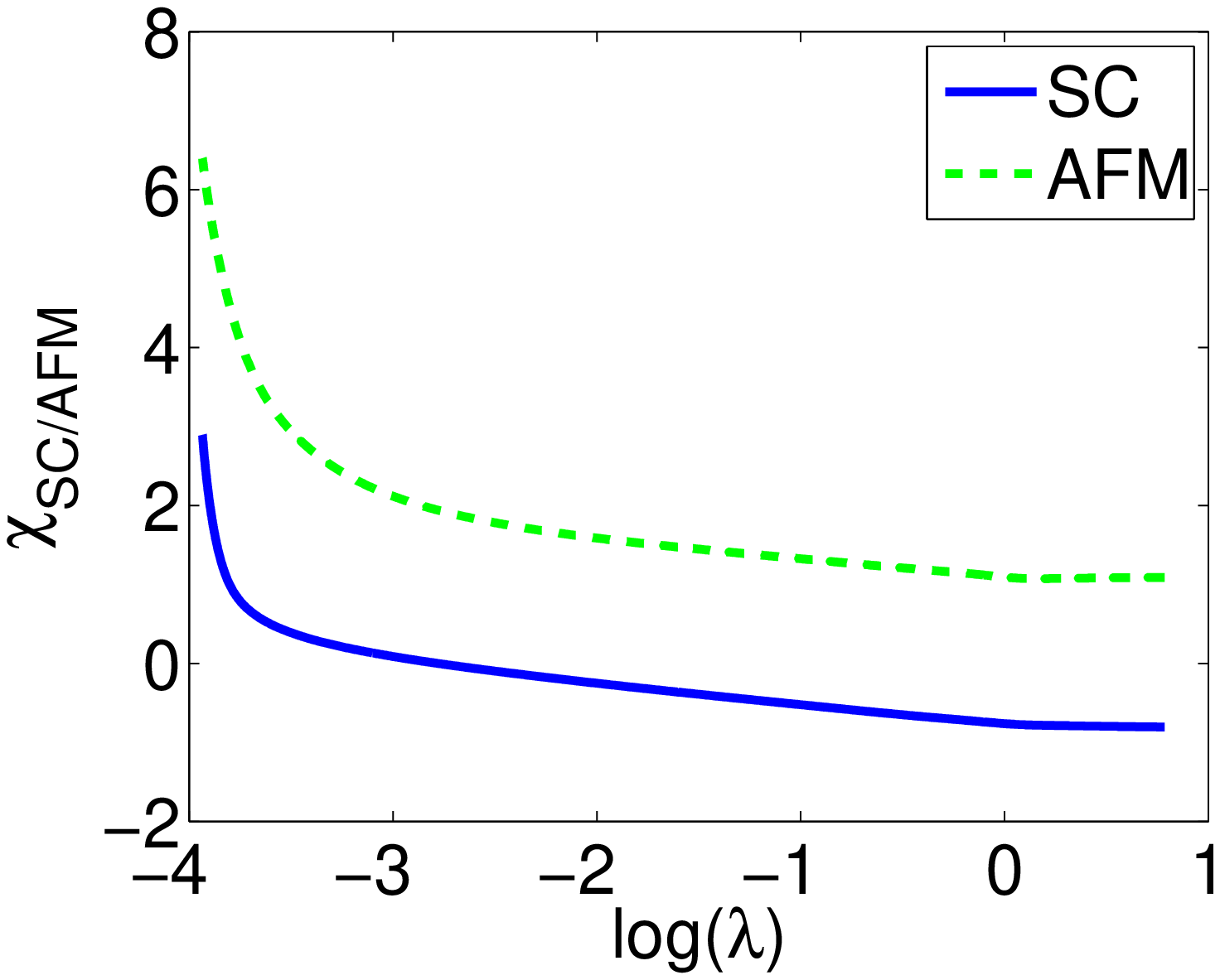}
\includegraphics[scale=0.23]{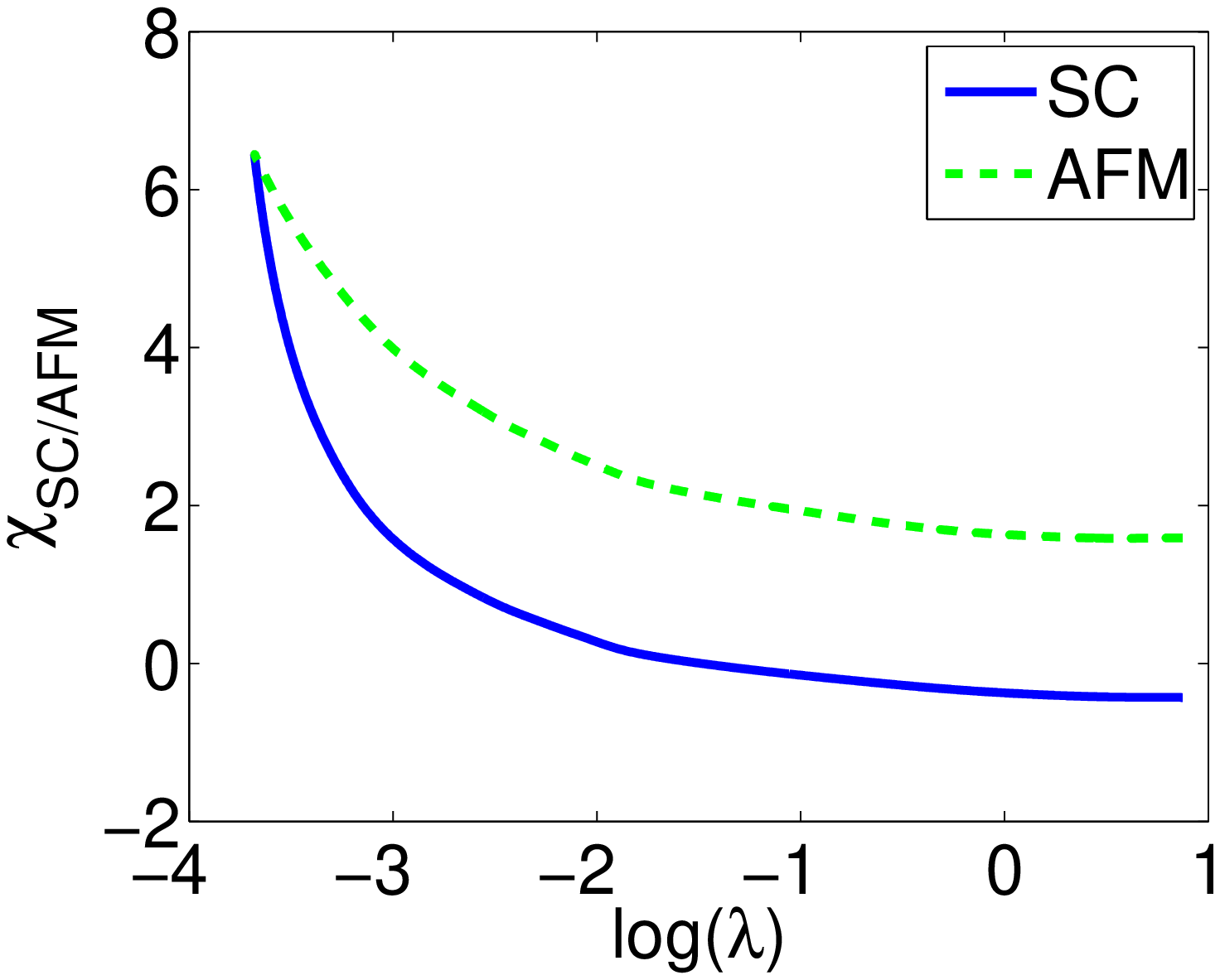}
\end{center}
\caption{Flow of the antiferromagnetic (green dashed line) and superconducting (blue solid line) interaction strengths for the La$_2$CuO$_4$-like band structure (left plot) and for the HgBa$_2$CuO$_4$-like band structure with $\Delta E \sim 2eV$ (right plot) for the same parameters as in Fig. \ref{3bandplot1}. In the first case, both channels compete, while in the second case, the $d$-wave pairing channel clearly dominates at low scales.}
\label{xiflow} 
\end{figure}   

We also compare the relative strength of the channels for different $\Delta E$ in Fig. \ref{xiflow}. For small $\Delta E$ in the left plot the AFM tendency is larger than the $d$-wave pairing strength, and both averages grow with similar exponent. As $\Delta E$ becomes larger, the AFM tendency is weakened due to poorer nesting of the Fermi surface. Then the superconducting channel is leading at low enough scales. 

Usually, the critical scale $\lambda_c$ is regarded as an upper estimate for actual transition temperatures into ordered states. Regarding the question what the true superconducting transition temperatures might be in the two situations compared here, it is  now very likely that in the case of strong competition between two channels, the transition temperature will actually be significantly reduced compared to $\lambda_c$, while for the cleaner pairing instability in the non-nested situation, $T_c$ might be closer to $\lambda_c$. We expect that an inclusion of the self-energy, which is not considered in this work, would capture this effect. Thus the true superconducting transition temperatures at small $\Delta E$ would be even smaller, and consequently increasing $\Delta E$ should increase the superconducting channel even more than in our calculation.

Summarizing these observations we state that the multiorbital model with orbital energy variations in accordance with actual material differences is able to reproduce to some degree the phenomenological tendency that critical temperatures can grow although the Fermi surface gets more rounded. 
This trend cannot be understood within the framework of one-band models. 
At least on a qualitative level, we have shown that the orbital mixing provides a mechanism for a $T_c$ enhancement at weak to moderate interactions. If this mechanism still works at the realistic interaction strength, it could, at least in part, be responsible for observed material trend for the $T_c$'s of the high $T_c$ cuprates.

\section{Conclusions}
We have studied two-orbital and three-orbital models on the two-dimensional square lattice that mimic the low-lying electronic structure of the high-$T_c$ cuprates. Using the fRG, we have computed the critical scale for $d$-wave pairing, which can be used as an estimate for the critical temperature for $d$-wave superconductivity. In one-band models with simple onsite repulsion, this energy scale is mainly dominated by the Fermi surface shape and decreases when the Fermi surface gets more rounded. The material trend for the real high-$T_c$ cuprates, or more precisely, the combination of experimental $T_c$ and electronic structure calculations for a series of cuprates, seems to contradict this trend, as materials with more rounded Fermi surfaces have higher experimental $T_c$'s. Our goal was to see if orbital admixture to the conduction band reverts the shape-related $T_c$ variation and allows one to understand this material trend. 

From our studies, we can draw two conclusions. First of all, for spin-fluctuation-induced $d$-wave pairing on the square lattice at a higher critical scale, it is best to have an energetically well-separated $d_{x^2-y^2}$-like band. We have shown that the admixture of bands with symmetry different from $d_{x^2-y^2}$ from above and below in energy typically reduces the critical scale for pairing compared to the situation without admixture. Reducing the consideration to the two-patch model allowed us to relate the orbital admixture to an additional repulsion in the $d$-wave channel, which explains the reduction of the $d$-wave pairing tendencies. In this setup it can also clearly seen that admixing orbitals of $s$, $d_{z^2}$, or, possibly,  $p_z$ character have a similar negative effect, as the sign structure of the admixture is the same.

With the knowledge that a single nearby band disturbs the pairing, we then analyzed situations with more than two bands. Here, a change of the orbital energies with one empty band moving closer toward the Fermi level can still result in a relative enhancement of the superconducting instability scale, consistent with the actual material trend. We have shown that in three-band models for the cuprates, the approach of the wider $s$ band can have a smaller negative effect on $T_c$ than the positive effect of the $d_{z^2}$ band moving further down to lower energies simultaneously. Hence we have identified a possible path how the $d$-wave $T_c$, generated by the spin-fluctuation mechanism, can be increased as a function of the energetic separation of orbitals near the Fermi level, although the Fermi surface of the $d_{x^2-y^2}$-dominated band gets more rounded. The main reason for this effect is that the $T_c$-raising reduction in the orbital admixture to the conduction band overcompensates the decrease in $T_c$ due to the rounder Fermi surface.

We have seen that choosing DFT-derived band-structure parameters we can qualitatively reproduce experimental trends in $T_c$ differences between La$_2$CuO$_4$ and HgBa$_2$CuO$_4$. However our calculations are only valid in the weak to moderate coupling regime, and if the interaction parameters are chosen to small the expected behavior is not rigorously reproduced. On the other hand, we have shown that the $T_c$ enhancement works better for larger interactions. The real cuprates are of course quite strongly correlated, beyond the interaction range where we can apply our method. There, the $T_c$ enhancement might be even more drastic, but based on our weak-coupling study, this is far from proven.
Furthermore, the $T_c$ increase may be even stronger than indicated by our numbers already for weaker interactions, as the less rounded Fermi surfaces have a stronger channel competition, which should result in a further lowering of the critical temperature for $d$-wave pairing.  Both these observations suggest that the mechanism considered in this work might be more effective in the actual materials. It should be interesting to look for the same tendency, e.g., with cluster-DMFT techniques. It should also be interesting to quantify the degree of admixture to the $d_{x^2-y^2}$ band through the series of high-$T_c$ cuprates that obeys the proclaimed relation between $T_c$ and the electronic structure, and to see whether the $T_c$ increase is correlated with this.

In this work, we have focused on one single aspect that may distinguish different cuprate materials, expressed by the orbital energies in three-orbital effective Hamiltonians. This allowed to obtain some understanding of how the pairing tendencies change. Very likely, there are other model parameters that exert additional influence on the energy scale of superconducting pairing. For example, depending on the  hybridization with the surrounding orbitals and the spread of the respective Wannier functions, the interaction parameters might vary. It appears to be an interesting topic to study these additional effects again in isolated form, and to understand their importance. Then it might be possible to compose a combined picture that might ultimately be used to guide the search for higher transition temperatures.

Acknowledgments: We acknowledge essential input by Ryotaro Arita, who provided the three-band model parameters. We further thank O.K. Andersen, E. Pavarini, M. Imada and W. Hanke for useful discussions. This work was supported by the DFG research unit FOR 1162 and the DFG priority program SPP1458.  

\bibliographystyle{plaindin}
\bibliography{bib}


%

 \end{document}